\documentclass[10pt, final, journal, letterpaper, twocolumn]{IEEEtran}

\usepackage{acronym}
\usepackage{amsfonts}
\usepackage{times}
\usepackage{cite}
\usepackage{amsmath}
\usepackage{array}
\usepackage{amssymb}
\usepackage{stfloats}
\usepackage{slashbox}
\usepackage{graphicx}
\usepackage{footnote}
\usepackage{amsthm}
\usepackage{booktabs}
\usepackage{array}
\usepackage{algorithmic}
\usepackage{algorithm}
\usepackage{subeqnarray}
\usepackage{cases}
\usepackage{threeparttable}
\usepackage{color}
\usepackage{epstopdf}
\usepackage{multirow}
\usepackage{tabularx}
\usepackage{enumerate}
\usepackage{multicol}
\usepackage{hyperref}
\usepackage{subfigure}

\renewcommand{\hat}{\widehat}
\renewcommand{\tilde}{\widetilde}

\newcommand{\secref}[1]{{Section}~\ref{#1}}
\newcommand{\figref}[1]{{Fig.}~\ref{#1}}

\def\bb0{{\mathbb{0}}}

\def\ba{{\mathbf{a}}}
\def\bb{{\mathbf{b}}}

\def\bp{{\mathbf{p}}}

\def\br{{\mathbf{r}}}
\def\bs{{\mathbf{s}}}
\def\bt{{\mathbf{t}}}
\def\bu{{\mathbf{u}}}
\def\bv{{\mathbf{v}}}
\def\bw{{\mathbf{w}}}
\def\bx{{\mathbf{x}}}
\def\by{{\mathbf{y}}}
\def\bz{{\mathbf{z}}}
\def\b0{{\mathbf{0}}}
\def\btheta{{\boldsymbol{\theta}}}
\def\bSigma{{\mathbf{\Sigma}}}

\def\bA{{\mathbf{A}}}
\def\bB{{\mathbf{B}}}
\def\bC{{\mathbf{C}}}

\def\bF{{\mathbf{F}}}
\def\bG{{\mathbf{G}}}
\def\bH{{\mathbf{H}}}
\def\bI{{\mathbf{I}}}
\def\bJ{{\mathbf{J}}}
\def\bK{{\mathbf{K}}}

\def\bM{{\mathbf{M}}}

\def\bR{{\mathbf{R}}}
\def\bS{{\mathbf{S}}}
\def\bT{{\mathbf{T}}}
\def\bU{{\mathbf{U}}}
\def\bV{{\mathbf{V}}}
\def\bW{{\mathbf{W}}}
\def\bX{{\mathbf{X}}}
\def\bY{{\mathbf{Y}}}
\def\bZ{{\mathbf{Z}}}

\def\sf0{{\mathsf{0}}}

\def\rm0{{\mathrm{0}}}

\def\Nt{{N_\mathrm{t}}}
\def\Nta{{N_\mathrm{t}^{\mathrm{a}}}}
\def\Nte{{N_\mathrm{t}^{\mathrm{e}}}}
\def\Nr{{N_\mathrm{r}}}
\def\Nra{{N_\mathrm{r}^{\mathrm{a}}}}
\def\Nre{{N_\mathrm{r}^{\mathrm{e}}}}
\def\Np{{N_\mathrm{p}}}
\def\Nb{{N_\mathrm{b}}}
\def\Nco{{N_\mathrm{co}}}
\def\Ncl{{N_\mathrm{cl}}}

\def\bwq{{\mathbf{w}_\mathrm{q}}}
\def\bwqk{{\mathbf{w}_{\mathrm{q},k}}}

\def\Pt{{P_{\mr{t}}}}

\def\Nx{{N_x}}
\def\Ny{{N_y}}

\newcommand{\mr}{\mathrm}
\def\j{\mathrm{j}}
\def\Re{\mathrm{Re}}
\def\Im{\mathrm{Im}}
\acrodef{CSI}[CSI]{channel state information}
\acrodef{CSIT}[CSIT]{channel state information at the transmitter}
\acrodef{CSIR}[CSIR]{channel state information at the receiver}
\acrodef{MIMO}[MIMO]{multiple-input multiple-output}
\acrodef{SISO}[SISO]{single-input single-output}
\acrodef{MISO}[MISO]{multiple-input single-output}
\acrodef{SIMO}[SIMO]{single-input multiple-output}
\acrodef{ADCs}[ADCs]{analog-to-digital convertors}
\acrodef{SNR}[SNR]{signal-to-noise ratio}
\acrodef{AWGN}[AWGN]{additive white Gaussian noise}
\acrodef{MRT}[MRT]{maximal ratio transmission}
\acrodef{DFT}[DFT]{Discrete Fourier Transform}
\acrodef{ULA}[ULA]{uniform linear array}
\acrodef{UPA}[UPA]{uniform planar array}
\acrodef{LS}[LS]{least squares}
\acrodef{ML}[ML]{maximum likelihood}
\acrodef{AML}[AML]{approximate ML}
\acrodef{LMMSE}[LMMSE]{linear MMSE}
\acrodef{ALMMSE}[ALMMSE]{approximate LMMSE}
\acrodef{QIHT}[QIHT]{quantized iterative hard thresholding}
\acrodef{QIST}[QIST]{quantized iterative soft thresholding}
\acrodef{PAPR}[PAPR]{peak-to-average power ratio}

\newcommand{\mc}[1]{\ensuremath{\mathcal{#1}}}

\newcommand{\Complex}{{\mathbb{C}}}

\newcommand{\tran}{^{\textsf{T}}}

\newcommand{\revision}[1]{\textcolor{black}{#1}}

\newcommand{\black}{\color{black}}
\begin{document}
	\title{Channel Estimation in Broadband Millimeter Wave MIMO Systems with Few-Bit ADCs}
	\author{\IEEEauthorblockN{Jianhua~Mo,~\IEEEmembership{Student~Member,~IEEE}, \IEEEauthorblockN{Philip~Schniter},~\IEEEmembership{Fellow,~IEEE}, and \IEEEauthorblockN{Robert W. Heath Jr.},~\IEEEmembership{Fellow,~IEEE}}\\
		\thanks{Jianhua Mo is with Samsung Research America, Richardson, TX 75082, USA (Email: mojianhua01@gmail.com)}
		\thanks{Robert W. Heath Jr. is with Wireless Networking and Communications Group, The University of Texas at Austin, Austin, TX 78712, USA (Email: \{rheath\}@utexas.edu). Their work was supported in part by the National Science Foundation under Grant CCF-1527079.}
		\thanks{Philip Schniter is with The Ohio State University, Columbus, OH 43210, USA (Email: \{schniter.1@osu.edu\}). His work was supported in part by the National Science Foundation under Grant CCF-1527162.}
		\thanks{The material in this paper was presented in part at the 2014 Asilomar Conference on Signals, Systems and Computers \cite{Mo_Jianhua_Asilomar14}.}
        }
	\maketitle	
\begin{abstract}
		We develop a broadband channel estimation algorithm for millimeter wave (mmWave) multiple input multiple output (MIMO) systems with few-bit analog-to-digital converters (ADCs). Our methodology exploits the joint sparsity of the mmWave MIMO channel in the angle and delay domains.  We formulate the estimation problem as a noisy quantized compressed-sensing problem and solve it using efficient approximate message passing (AMP) algorithms.  In particular, we model the angle-delay coefficients using a Bernoulli-Gaussian-mixture distribution with unknown parameters and use the expectation-maximization (EM) forms of the generalized AMP (GAMP) and vector AMP (VAMP) algorithms to simultaneously learn the distributional parameters and compute approximately minimum mean-squared error (MSE) estimates of the channel coefficients.
                We design a training sequence that allows fast, FFT-based implementation of these algorithms while minimizing peak-to-average power ratio at the transmitter, making our methods scale efficiently to large numbers of antenna elements and delays.
                We present the results of a detailed simulation study that compares our algorithms to several benchmarks.  Our study investigates the effect of SNR, training length, training type, ADC resolution, and runtime on channel estimation MSE, mutual information, and achievable rate.
                \revision{It shows that, in a mmWave MIMO system, the methods we propose to exploit joint angle-delay sparsity allow one-bit ADCs to perform comparably to infinite-bit ADCs at low SNR, and 4-bit ADCs to perform comparably to infinite-bit ADCs at medium SNR.}
\end{abstract}

\begin{IEEEkeywords}
	Low resolution analog-to-digital converter, millimeter wave, channel estimation, approximate message passing
\end{IEEEkeywords}
	
\section{Introduction}
                Millimeter wave (mmWave) communication is a promising technology for future outdoor cellular systems due to its potential to use very high bandwidth channels \cite{Rappaport_Book15}.
                But larger bandwidths place difficult demands on the receiver's analog-to-digital converters (ADCs). For example, at rates above $100$ Msamples per second, ADC power consumption increases quadratically with sampling frequency\cite{Murmann_FTFC13}.  High-precision ADCs (e.g., $\geq6$ bits) with bandwidths sufficient for mmWave systems (e.g., $\geq1$ Gsamples/s) are either unavailable or may be too costly and power-hungry for portable devices\cite{Le_Bin_SPM05}. One possible solution is to employ low-resolution ADCs, which enjoy low power consumption and simple hardware implementations. Low resolution ADCs have further benefits in MIMO systems, where a large number of ADCs are needed when digital baseband processing of all antenna outputs is desired. For example, in massive MIMO systems, it has been suggested to equip the base station with dozens of antennas and the same number of 1-bit ADCs%
                \revision{\cite{Jacobsson_ICC15,Wen_Chao-Kai_TSP16,Choi_TCOM16}}.

                In this paper, we consider a receiver architecture based on few-bit (i.e., 1-4 bit) ADCs, which act to quantize the inphase and quadrature baseband received signals. The achievable rate of the quantized MIMO channel was studied in \cite{Nossek_IWCMC06, Mezghani_ISIT07, Mezghani_ISIT12} assuming channel-state information at the receiver (CSIR) was perfect but channel-state information at the transmitter (CSIT) was absent. In the latter case, and assuming equal transmission power at each antenna, \cite{Mezghani_ISIT07} showed that QPSK is the optimum signaling strategy at low SNR. The perfect-CSIT case was studied in our previous work \cite{ Mo_Jianhua_TSP15}, where constellation design methods were proposed to maximize the achievable rate. These methods achieve much higher rates than QPSK signaling, especially at high SNR, but require CSIT.

                Due to the nonlinear nature of quantization, channel estimation with few-bit ADCs is challenging. To estimate broadband SISO channel coefficients, the work \cite{Dabeer_ICC10, Zeitler_TSP12} proposed to transmit periodic bursty training sequences, dither the ADCs, and estimate each tap separately.  MIMO channel estimation is even more challenging because the linear \emph{combination} of transmitted signals from different antennas is quantized. In \cite{Risi_arxiv14} and \cite{Wang_Shengchu_TWC15}, the MIMO channel was estimated using \ac{LS} methods. In particular, the quantization error was treated as additive white Gaussian noise. As a result, a large estimation error was introduced.

                To take into account quantization effects, iterative channel estimation methods using Expectation Maximization (EM) were proposed and analyzed in \cite{Ivrlac_WSA07, Mezghani_WSA10,Mo_Jianhua_Asilomar14}. The proposed methods have high complexity, however, since each EM iteration computes a matrix inverse and many iterations are needed for convergence. In addition, \cite{Ivrlac_WSA07, Mezghani_WSA10} considered the MIMO channel with small antenna arrays in the lower frequency UHF (ultra high frequency) band, and thus did not take into account the sparsity of mmWave channels \cite{Rappaport_TCOM15,WP_5G_Channel_Model,Mo_Jianhua_Asilomar14,Schniter_Asilomar14}.

                In recent work \cite{Wen_Chao-Kai_TSP16,Wang_Hanqing_ICCS16,Zhang_Ti-Cao_TWC16,Stockle_SPAWC16}, Approximate Message Passing (AMP) algorithms were used for channel estimation and (or) symbol detection in the massive MIMO channel with few-bit ADCs. These works assumed that the channel coefficients follow an IID Gaussian distribution and did not exploit the sparsity inherent in mmWave channels. (Delay-domain sparsity was exploited in \cite{Stockle_SPAWC16} but not angle-domain sparsity.) Also, since they were based on the massive-MIMO channel, these works did not exploit structures present in the broadband mmWave model that can be exploited for significant complexity reduction.

                In this paper, we propose high-performance AMP-based channel estimation schemes for broadband mmWave MIMO channels with few-bit ADCs.  The main contributions of our work are summarized as follows.
                \begin{itemize}
                        \item 
                            \revision{We formulate the problem of estimating broadband mmWave channels under few-bit ADCs as a noisy, quantized, compressed-sensing problem.
                            By leveraging sparsity in both the angle and delay domains, the massive MIMO channel can be accurately estimated by efficient algorithms and with relatively short training sequences. To our knowledge, our work is the first to take this approach.}
                        \item For channel estimation, we consider two AMP algorithms: the Generalized AMP (GAMP) algorithm from \cite{Rangan_ISIT11} and the Vector AMP (VAMP) algorithm from \cite{Rangan_arxiv16,Schniter_Asilomar16}. For both, we use EM extensions \cite{Vila_TSP13,Fletcher_ICASSP17} to avoid the need to specify a detailed prior on the channel distribution. \revision{The GAMP and VAMP algorithms provide nearly minimum-MSE estimates with relatively low complexity in large-scale estimation problems and are therefore suitable for estimating broadband massive MIMO channels.}
                        \item We propose a novel training sequence design that results in low channel-estimation error, low complexity, and low peak-to-average power ratio.  Low complexity is achieved through a novel FFT-based implementation that will be described in the sequel.
                        \item We undertake a detailed experimental study of important design choices, such as
                        ADC precision,
                        the type and length of training sequence, and
                        the type of estimation algorithm.
                        When evaluating algorithms, we consider both performance and complexity.  For performance, we consider several metrics: mean-square error (MSE), mutual information, and achievable rate.
                \end{itemize}	
        From our experimental study, our main findings are as follows.
                \begin{itemize}
                        \item Compared to other algorithms of which we are aware, the EM-VAMP algorithm has a superior performance-complexity tradeoff.
                        \item Our FFT-based implementation facilitates low-complexity estimation of channels with large antenna numbers (e.g., $64 \times 64$) and delay spreads (e.g., $16$ symbol intervals).
                        \item Relative to infinite-bit ADCs, $1$-bit ADCs incur only small performance losses at low SNR, and $3$-$4$-bit ADCs incur only small losses up to medium SNRs.
                        \item The MSEs of the EM-GAMP and EM-VAMP algorithms decay exponentially with training length, and the achievable rate is usually maximized by sending a relatively short training sequence.
                \end{itemize}

        A simple version of our proposed methodology was first published in \cite{Mo_Jianhua_Asilomar14}.  Relative to \cite{Mo_Jianhua_Asilomar14}, our current work
        expands from one-bit to few-bit ADCs,
        expands from narrowband to broadband channels,
        considers leakage effects in sparse-channel modeling,
        considers four training sequence designs instead of one,
        considers the VAMP algorithm as well as the GAMP algorithm,
        proposes an FFT-based implementation that facilitates many more antennas and delays,
        considers Gaussian-mixture (GM) as well as Bernoulli-Gaussian priors,
        and incorporates EM learning of the prior parameters.
        Also, relative to \cite{Mo_Jianhua_Asilomar14}, the experimental study in our current work is much more elaborate.

                The paper is organized as follows. In Section II, we describe the broadband MIMO system model with few-bit ADCs. The channel characteristics of mmWave communications are then summarized in Section III. In Section IV, we present our channel-estimation algorithms and training-sequence designs. Simulation results are presented in Section V, followed by conclusions in Section VI.

                \emph{Notation} : $a$ is a scalar, $\ba$ is a vector and $\bA$ is a matrix.
                $\mr{tr}(\bA)$, $\bA\tran$, $\bA^*$ and $||\bA||_F$ represent the trace, transpose, conjugate transpose and Frobenius norm of a matrix $\bA$.
                $\bA \otimes \bB$ denotes the Kronecker product of $\bA$ and $\bB$.
                $\mr{vec}(\bA)$ is a vector stacking all the columns of $\bA$.
                $\mr{unvec}(\cdot)$ is the opposite operation of $\mr{vec}(\cdot)$.
                $\mathcal{CN}(x;\mu, \sigma^2) \triangleq \frac{1}{\pi \sigma^2} \exp (- |x-\mu|^2/\sigma^2)$ is the probability density function of the circularly symmetric complex-Gaussian distribution  with mean $\mu$ and variance $\sigma^2$.
                Finally,
                $\j\triangleq\sqrt{-1}$.

                \section{Broadband MIMO with Few-bit ADC}
                \begin{figure}[t]
                        \begin{centering}
                                \includegraphics[width=1\columnwidth]{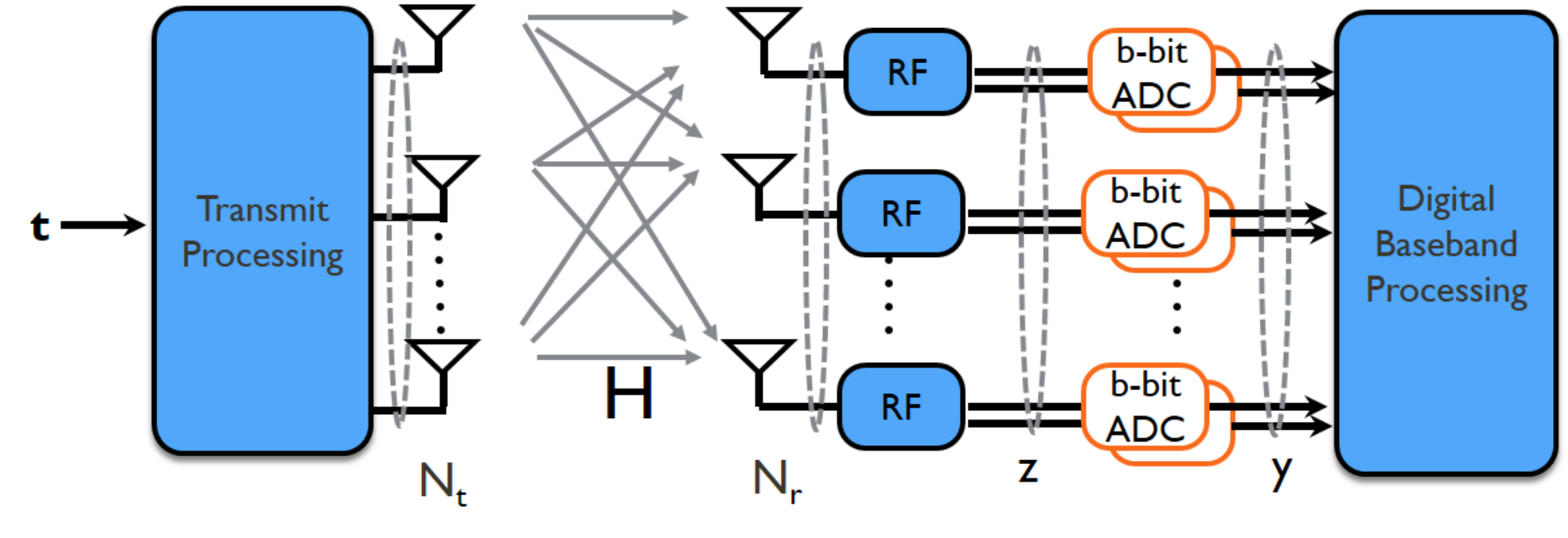}
                                \vspace{-0.7cm}
                                \centering
                                \caption{A $\Nr \times \Nt$ MIMO system with one-bit quantization at the receiver.  For each receiver antenna, there are two few-bit ADCs. Note that there is no limitation on the structure of the transmitter. }\label{fig:System_model}
                        \end{centering}
                        \vspace{-0.3cm}
                \end{figure}

                We consider a MIMO system with few-bit ADCs, as illustrated in Fig. \ref{fig:System_model}.
                The transmitter is equipped with $N_{\mr{t}}$ antennas and the receiver is equipped with $N_{\mr{r}}$ antennas.
                A total of $2\Nr$ few-bit ADCs separately quantize the real and imaginary parts of the received signal of each antenna.
                Assuming that the delay spread of the channel is limited to $L$ symbol intervals and that carrier and symbol synchronization have already been performed, the quantizer output $\by[i]\in \mathbb{C}^{\Nr \times 1}$ at time $i$ can be written as
                \begin{equation} \label{eq:y[i]}
                \by[i] =  \mathcal{Q} \left(\sum_{\ell=0}^{L-1} \bH[\ell] \bt[i-\ell] + \bw[i] \right),
                \end{equation}
                where $\bH[\ell] \in \mathbb{C}^{\Nr \times \Nt}$ is the baseband channel impulse response at lag $\ell$, $\bt[i] \in \mathbb{C}^{\Nt \times 1}$ is the transmitted symbol at time $i$ with average transmit power $\mathbb{E}\left [\bt[i]^* \bt[i]\right] = \Pt$, and $\bw[i] \sim \mathcal{CN}(\mathbf{0}, \sigma_w^2\bI_{\Nr})$ is additive white Gaussian noise.
                Furthermore, $\mathcal{Q}(\cdot)$ denotes the quantization function, which is applied component-wise and separately to the real and imaginary parts.

                In this paper, we assume that uniform mid-rise quantization is used.
                In particular, if $x$ is a complex-valued scalar $x$, then $y=\mathcal{Q}(x)$ means
                \begin{align}\label{eq:few-bit_ADC}
                        y &=  \mr{sign} \left( \Re(x) \right)  \left( \min \left( \left\lceil \frac{|\Re(x)|}{\Delta_{\Re}} \right \rceil, 2^{b-1} \right)  - \frac{1}{2} \right) \Delta_{\Re} \nonumber \\
                        &\quad + \j \; \mr{sign} \left( \Im(x) \right)  \left( \min \left( \left\lceil \frac{|\Im(x)|}{\Delta_{\Im}} \right \rceil, 2^{b-1} \right)  - \frac{1}{2} \right) \Delta_{\Im},
                \end{align}
                where $\Delta_{\Re} \triangleq \left(\mathbb{E} \left[ |\Re(x)|^2 \right]\right)^{\frac{1}{2}} \Delta_b$ and $\Delta_{\Im} \triangleq \left(\mathbb{E} \left[ |\Im(x)|^2 \right]\right)^{\frac{1}{2}} \Delta_b$,
                and where $\Delta_b$ is a stepsize that will be discussed in the sequel.
                In the special case of one-bit quantization, \eqref{eq:few-bit_ADC} becomes
                \begin{align}\label{eq:one-bit_ADC}
                        y &=  \mr{sign} \left( \Re(x) \right) \sqrt{\frac{2}{\pi}} \left( \mathbb{E} \left[ |\Re(x)|^2 \right] \right)^{\frac{1}{2}} \nonumber \\
                        &\quad + \j \; \mr{sign} \left( \Im(x) \right) \sqrt{\frac{2}{\pi}} \left( \mathbb{E} \left[ |\Im(x)|^2 \right] \right)^{\frac{1}{2}}.
                \end{align}
                The average powers $\mathbb{E} \left[ |\Re(x)|^2 \right]$ and $\mathbb{E} \left[ |\Im(x)|^2 \right]$ can be easily measured by analog circuits before the ADC, as in automatic gain control (AGC).
                In this paper, $x$ is circularly symmetric, and so $\mathbb{E} \left[ |\Re(x)|^2 \right]= \mathbb{E} \left[ |\Im(x)|^2 \right] = \frac{1}{2} \mathbb{E} \left[ |x|^2 \right]$, implying that $\Delta_{\mr{Re}} = \Delta_{\mr{Im}}$.

                The quantization stepsize $\Delta_b$ is usually chosen to minimize the quantization MSE assuming a Gaussian input signal (see, e.g., \cite{Max_IRE60}).
                These values of $\Delta_b$ are given in Table~\ref{tab:Delta} assuming a unit-power input. The normalized MSE (NMSE), defined as $\eta_b\triangleq \mathbb{E} \big[ \left| \mathcal{Q}(x) -  x \right|^2 \big]/\mathbb{E} \big[ \left| x \right| ^2 \big]$, and the signal to quantization noise ratio (SQNR), defined as $10 \log_{10} \frac{1}{\eta_b}$, are also listed in Table~\ref{tab:Delta}.
                From the table it can be seen that $\Delta_b \sim 2^{-b}$, $\eta_b \sim 2^{-2b}$ and $\text{SQNR}\approx 5b$ dB.
                Notice that 4-bit ADCs yield quantization noise power around 20 dB below the signal power, which suggests that increasing ADC resolution beyond 4 bits should yield negligible performance improvement in the low and medium SNR regimes. This intuition will be verified by our simulations in Section \ref{sec:simulation}.
                \begin{table*}
                        \centering
                        \caption{The optimum uniform quantizer for a Gaussian unit-variance input signal \cite{Max_IRE60} }
                        \label{tab:Delta}
                        \begin{tabular}{|c|c|c|c|c|c|c|c|c|}
                                \hline
                                Resolution $b$  & 1-bit  & 2-bit  & 3-bit & 4-bit & 5-bit & 6-bit & 7-bit & 8-bit\\
                                \hline
                                Stepsize $\Delta_b$  & $\sqrt{\frac{8}{\pi}}$ ($\approx$ 1.5958) & 0.9957 & 0.586 & 0.3352 & 0.1881 & 0.1041 & 0.0569 & 0.0308 \\
                                \hline
                                NMSE $\eta_b$ & $\frac{2-\pi}{\pi}$ ($\approx$ 0.3634) & 0.1188 & 0.03744 & 0.01154 & 0.003504 & 0.001035 & 0.0002999 & 0.00008543\\
                                \hline
                                SQNR (dB) & $10 \log_{10} \frac{\pi}{2-\pi}$ ($\approx$ 4.4) & 9.25 & 14.27 & 19.38 & 24.55 & 29.85 & 35.23 & 40.68\\
                                \hline
                        \end{tabular}
                \end{table*}


                \section{Sparsity of the mmWave Channel Model}
                In this section, we present the mmWave channel model assumed in our paper. The characteristics of the mmWave channel, especially the sparsity in the angle-delay domain, will be exploited by our proposed channel estimation algorithm.

                \subsection{Clustered MIMO Channel Model}
                The mmWave channel can be modeled using $\Ncl$ multipath clusters, where the $n$th cluster comprises $N_{\mr{path}}^n$ paths \cite{WP_5G_Channel_Model}.
                For the $m$th path of the $n$th cluster, we use $\alpha_{n, m}$, $\tau_{n, m}$, $\varphi^{\mr{r}}_{n, m}(\text{or}\,\theta^{\mr{r}}_{n, m})$, $\varphi^{\mr{t}}_{n, m}(\text{or}\,\theta^{\mr{t}}_{n, m})$ to denote the complex gain, delay, azimuth (or zenith) angle of arrival, and azimuth (or zenith) angle of departure, respectively.
                Using these quantities, the channel impulse response from \eqref{eq:y[i]} can be written as
                \begin{align} \label{eq:Cluster_channel}
                \bH[\ell] &= \sum_{n=1}^{\Ncl} \sum_{m=1}^{N_{\mr{path}}^n} \alpha_{n, m} \ba_{\mr{r}}(\varphi^{\mr{r}}_{n, m}, \theta^{\mr{r}}_{n, m}) \ba_{\mr{t}}^*(\varphi^{\mr{t}}_{n, m}, \theta^{\mr{t}}_{n, m}) \nonumber \\
                &\quad \times p( \ell T - \tau_{n, m}) , \quad 0 \leq \ell < L,
                \end{align}
                where $p(t)$ includes the effects of pulse shaping and analog/digital filtering, and where $\ba_{\mr{r}}$ and $\ba_{\mr{t}}$ are the array response vectors of the receive and transmit antenna arrays, respectively.

                We assume that the receive and transmit arrays are each configured as a \ac{UPA}. This assumption is quite common.
                \revision{For example, in the massive MIMO systems described in \cite{Chen_Xiao-Ping_TAP10, Guidi_TMC16}, UPAs with more than 100 antennas were tested.}
                In the mmWave 5G cellular system prototype described in \cite{Hong_Wonbin_COMM14}, there is a 256-element ($16 \times 16$) UPA array panel at the base station, and two sets of $1\times 16$ UPA arrays in the top and bottom portions of the mobile phone.

                \subsection{Angle-Delay Representation}

                The MIMO channel coefficients $\bH[\ell]\in\Complex^{\Nr\times\Nt}$ are expressed in what is known as the ``antenna aperture domain.''
                We find it convenient to instead work with ``angle domain'' coefficients $\bX[\ell]\in\Complex^{\Nr\times\Nt}$, as proposed in \cite{Sayeed_TSP02}.
                The two representations are connected through
                \begin{align} \label{eq:Virtual_channel}
                \bH[\ell] = \bB_{\Nr} \bX[\ell] \bB_{\Nt}^*,
                \quad 0\leq \ell < L ,
                \end{align}
                where $\bB_{\Nr}\in\Complex^{\Nr\times\Nr}$ and $\bB_{\Nt}\in\Complex^{\Nt\times\Nt}$ are the steering matrices for the transmitter and receiver arrays, respectively. \revision{Denote $\Nre$ ($\Nra$) as the number of receive antennas in the elevation (azimuth) direction, and $\Nte$ ($\Nta$) as the number of transmit antennas in the elevation (azimuth) direction.}
                With an $\Nre \times \Nra$ receive \ac{UPA} and an $\Nte\times\Nta$ transmit \ac{UPA}, we have that $\Nr = \Nra \Nre$, $\Nt = \Nta \Nte$, and
                \cite{Brady_SPAWC14}
                \begin{align} \label{eq:UPA_virtual_channel}
                \bH[\ell] = \left( \bF_\Nra \otimes \bF_\Nre \right) \bX[\ell] \left(\bF_\Nta \otimes \bF_\Nte \right)^* ,
                \end{align}
                where $\bF_{\Nra} \in \mathbb{C}^{\Nra \times \Nra}$, $\bF_{\Nre} \in \mathbb{C}^{\Nre \times \Nre}$, $\bF_{\Nta} \in \mathbb{C}^{\Nta \times \Nta}$, and $\bF_{\Nte} \in \mathbb{C}^{\Nte \times \Nte}$ are unitary \ac{DFT} matrices.
                The $(i,j)$th entry of the matrix $\bX[\ell]$ can be interpreted as the channel gain between the $j$th discrete transmit angle and the $i$th discrete receive angle \cite{Sayeed_TSP02}.

                According to the measurement results reported in \cite{WP_5G_Channel_Model}, the number of clusters $\Ncl$ tends to be relatively few in the mmWave band as compared to lower-frequency bands. Also, the number of antenna elements used tends to be large in order to counteract the effects of path loss, which is much more severe in the mmWave band as compared to low-frequency bands.  Hence, in mmWave applications, the number of channel clusters is usually much fewer than the number of scalar coefficients in $\left\{\bH[\ell]\right\}_{l=0}^{L-1}$ or $\left\{\bX[\ell]\right\}_{l=0}^{L-1}$, i.e., $\Ncl \ll \Nr \Nt L$.

\begin{figure}
                \centering
                \subfigure[Antenna domain channel]{
                        \includegraphics[width=0.8\columnwidth]{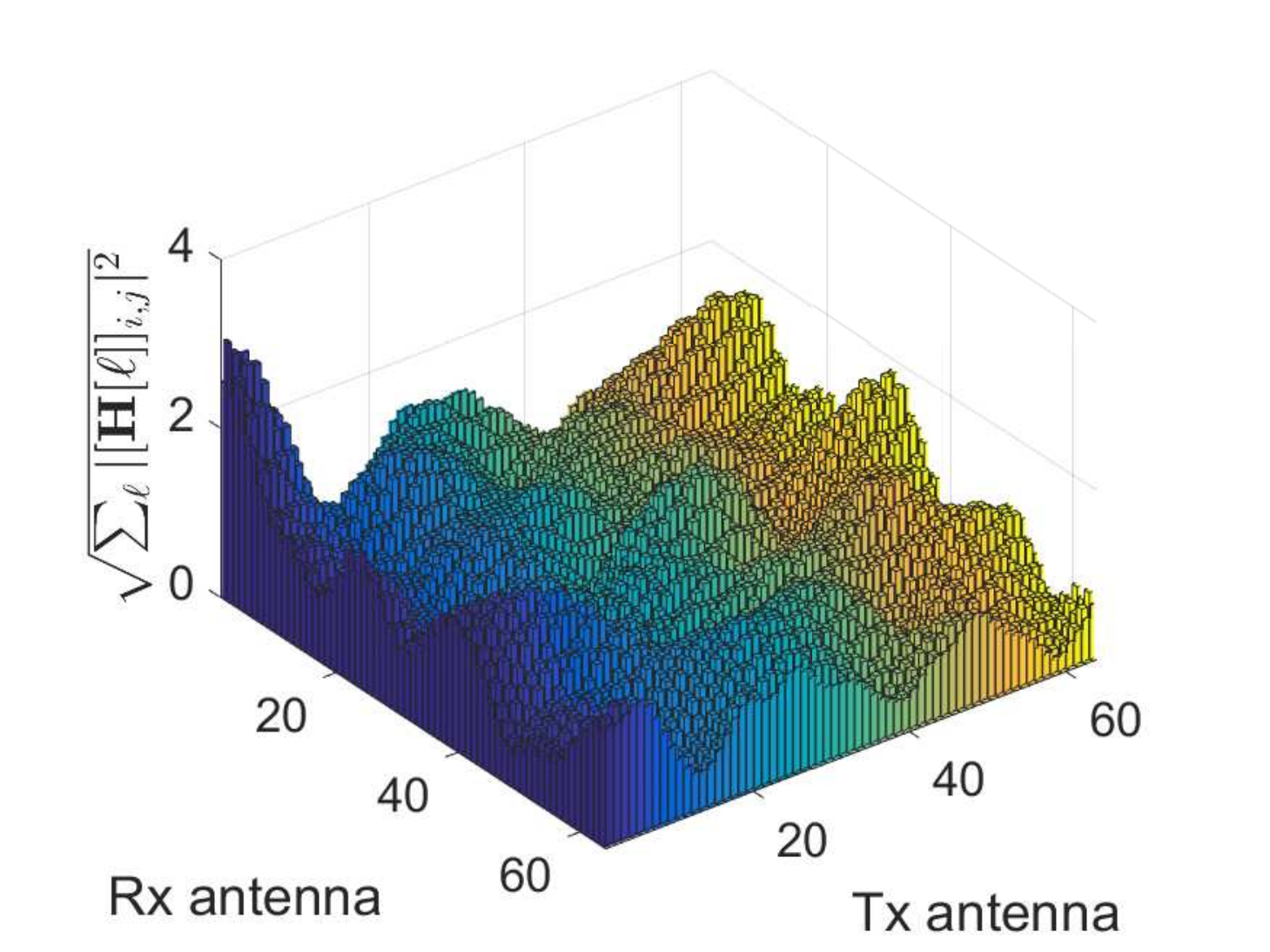}
                        \label{fig:Spatial_Domain_Channel}
                }
                \subfigure[Angle domain channel]{
                        \includegraphics[width=0.8\columnwidth]{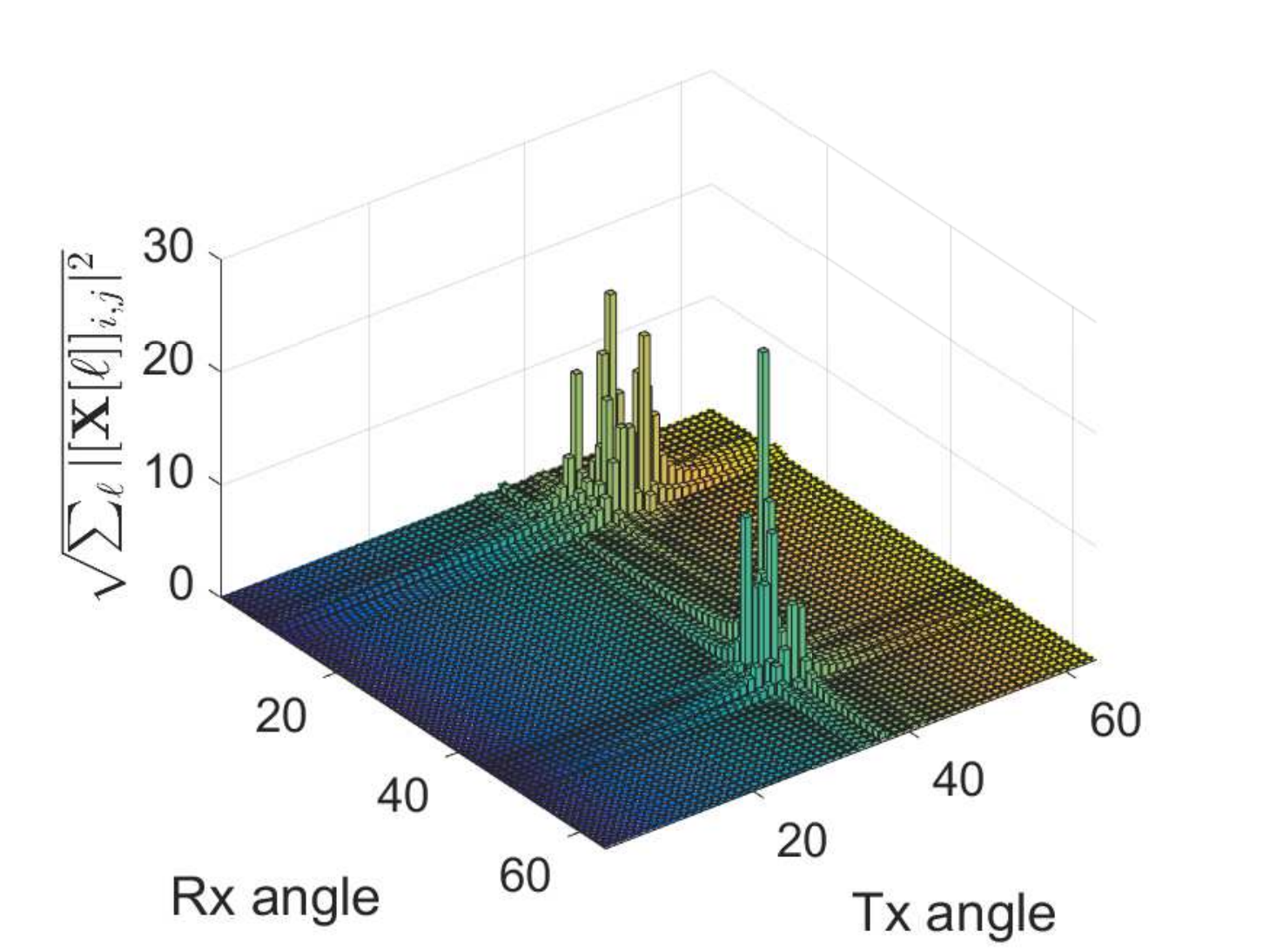}
                        \label{fig:Angular_Domain_Channel}
                }
                \caption{An example of a broadband MIMO channel with $\Ncl=2$ multipath clusters, as seen through a link with $8\times 8$ UPAs at both ends.  Subfigure (a) plots \revision{in linear scale} the mean-squared coefficient magnitude in the antenna aperture domain $\sqrt{\sum_{\ell} |[\bH[\ell]]_{i,j}|^2}$, and subfigure (b) plots \revision{in linear scale} the same for the angle domain $\sqrt{\sum_{\ell}|[\bX[\ell]]_{i,j}|^2}$.  The angle domain representation in (b) clearly shows two groups of large-magnitude entries, each of which corresponds to one multipath cluster.}\label{fig:Beam_sparse}
        \end{figure}

\revision{To show precisely how channel sparsity manifests in the angle domain,} \figref{fig:Beam_sparse} provides an example of a \revision{simulated} broadband mmWave channel with delay spread $L=16$ and $\Ncl=2$ multipath clusters, each consisting of $10$ paths and $7.5$ degrees of azimuth and elevation angular spread, as seen through $8 \times 8$ UPAs at both ends of the link (i.e., $\Nt=\Nr=64$) and raised-cosine pulse-shape filtering with roll-off factor $=0$. \revision{These parameters follow the urban macro (UMa) NLOS channel measurements at 28 GHz reported in \cite{WP_5G_Channel_Model}.}
        \figref{fig:Spatial_Domain_Channel} plots the mean-squared coefficient magnitude in the antenna aperture domain, i.e., $\sqrt{\sum_{\ell} |[\bH[\ell]]_{i,j}|^2}$, while subfigure (b) plots the mean-squared coefficient magnitude in the angle domain, i.e., $\sqrt{\sum_{\ell}|[\bX[\ell]]_{i,j}|^2}$.
        \figref{fig:Angular_Domain_Channel} shows that, in the angle domain, the channel energy is concentrated in two locations, each corresponding to one multipath cluster.

        We emphasize that the angle-delay channel $\{\bX[\ell]\}_{\ell=0}^{L-1}$ is \emph{jointly} sparse in the angle and delay domains.  This joint sparsity is illustrated in \figref{fig:True_Channel} for a $4\times 16$ MIMO channel with delay spread $L=16$ (see the details in \secref{sec:simulation}).  However, $\{\bX[\ell]\}_{\ell=0}^{L-1}$ is \emph{not exactly sparse}, exhibiting what is known as leakage, in that none of its coefficients are expected to be exactly zero valued.  The approximate nature of the sparsity is also visible in \figref{fig:True_Channel}. Our approach based on GAMP will be robust to leakage effects.

	   \section{Proposed Channel Estimation Algorithm}
	   Motivated by the channel model given in the last section, we now develop an efficient algorithm to estimate the approximately sparse angle-delay domain channel from few-bit measurements and a known training sequence.
	
	   \subsection{Problem Formulation}
	
	   We assume that the training consists of a block transmission of length $\Np$, denoted as $\bT \in \mathbb{C}^{\Nt \times \Np} $, with a cyclic prefix of length $L$.
	   After discarding the cyclic prefix, the measurements take the form
	   \begin{align}
	   \bY=\mathcal{Q}(\bZ+\bW),
	   \end{align}
	   where $\bW$ is additive Gaussian noise and $\bZ\in\Complex^{\Nr\times\Np}$ is
	   the unquantized noiseless received signal block, of the form
	   \begin{align}
	   \bZ
	   &= \sum_{\ell=0}^{L-1} \bH[\ell] \bT \bJ_\ell \\
	   &= \sum_{\ell=0}^{L-1} \bB_{\Nr} \bX[\ell] \bB^*_{\Nt} \bT \bJ_\ell \\
	   &= \bB_{\Nr}
	   \underbrace{\begin{bmatrix}\bX[0]& \bX[1]& \bX[2]& \cdots& \bX[L-1]\end{bmatrix}}_{\textstyle \triangleq\bX} \nonumber \\
	   & \quad \times
	   \underbrace{ \begin{bmatrix}
	   \bB^*_{\Nt}\\
	   &\bB^* _{\Nt}\\
	   &&\ddots\\
	   &&& \bB^*_{\Nt}
	   \end{bmatrix}}_{\textstyle =\bI_L \otimes \bB^*_{\Nt}}
	   \underbrace{ \begin{bmatrix}
	   \bT \bJ_0 \\
	   \bT \bJ_1 \\
	   \vdots\\
	   \bT \bJ_{L-1}
	   \end{bmatrix}}_{\textstyle \triangleq \tilde{\bT} } \label{eq:tilde_T}\\
	   &= \bB_{\Nr} \bX
	   \underbrace{ \begin{bmatrix}
	           \bB^*_{\Nt} \bT \bJ_0 \\
	           \bB^*_{\Nt} \bT \bJ_1 \\
	           \vdots\\
	           \bB^*_{\Nt} \bT \bJ_{L-1}
	           \end{bmatrix}}_{\textstyle \triangleq \bC },
	   \label{eq:output_z}
	   \end{align}
	   where $\bJ_\ell \in \mathbb{R}^{\Np \times \Np}$ is the $\ell$-circulant-delay matrix.
	
	   We now rewrite $\bZ$ in vector form $\bz \triangleq  \mr{vec}(\bZ)$ as
	   \begin{align}
	   \bz
	   &= \left(\bC\tran \otimes \bB_{\Nr} \right) \mr{vec}(\bX).
	   \end{align}
	   Defining $\by \triangleq \mr{vec}(\bY)$, $\bx \triangleq \mr{vec}(\bX)$, and $\bw \triangleq \mr{vec}(\bW)$,
	   the quantized noisy output becomes
	   \begin{align} \label{eq:1bit_CV}
	   \by= \mathcal{Q} \Bigg( \underbrace{\left( \bC\tran \otimes \bB_{\Nr} \right)}_{\textstyle \triangleq\bA} \bx  + \bw \Bigg).
	   \end{align}
	   The channel-estimation problem reduces to the following: estimate the angle-delay channel coefficients $\bx \in \mathbb{C}^{\Nt \Nr L}$ from the noisy quantized received signal $\by\in \mathbb{C}^{\Nr \Np \times 1}$ under the known linear transform $\bA\in \mathbb{C}^{\Nr \Np \times \Nt \Nr L}$.
	   For use in the sequel, we define $\Ny \triangleq \Nr \Np$ and $\Nx \triangleq \Nt \Nr L$.

	   Because the angle-delay channel is sparse (approximately due to leakage), this channel estimation problem can be viewed as an instance of \emph{noisy quantized compressed sensing}.
	   Several methods have been proposed for noisy quantized compressed sensing, including
	   quantized iterative hard thresholding (QIHT) \cite{Jacques_IT13,Jacques_arxiv13},
	   convex relaxation \cite{Zymnis_SPL10,Plan_IT13},
	   and GAMP \cite{Kamilov_TSP12,Mezghani_WSA12}.
	   It has also been proposed to treat the quantization error as if it were additive white Gaussian noise \cite{Risi_arxiv14,Wang_Shengchu_TWC15}.
	   All of these methods are strongly dependent on the sparsity rate assumed for the channel, either explicitly or through the specification of a regularization term or prior distribution.   \revision{In practice, though, the the channel sparsity rate, $1-\lambda_0$ in (17) and (18), is a priori unknown.}
	
	   \subsection{EM-AMP Algorithms}
	   For channel estimation, we propose to \revision{modify} two approaches that combine expectation maximization (EM) with AMP as a means of avoiding the need to specify a prior.  The first is based on a combination of generalized AMP (GAMP) \cite{Rangan_ISIT11} and EM, which was proposed in \cite{Vila_TSP13} but, to our knowledge, has never been applied to noisy few-bit quantized compressive sensing.  The second is based on a combination of the recently proposed vector AMP (VAMP) \cite{Rangan_arxiv16, Schniter_Asilomar16} and EM \cite{Fletcher_ICASSP17}.  To our knowledge, neither VAMP nor EM-VAMP have been applied to noisy few-bit quantized compressive sensing.
	   We focus on these AMP approaches because they offer nearly minimum MSE (MMSE) performance while being computationally efficient.
	   We provide background on these methods \revision{here to make our paper self-contained}.
	
	   \subsubsection{EM-GAMP}
	   Suppose that a random vector $\bx$ with IID components $x_i\sim p_X$ is linearly transformed to produce $\bz=\bA\bx$, which then propagates through a probabilistic measurement channel $p(\by|\bz)=\prod_i p_{Y|Z}(y_i|z_i)$.  Given knowledge of $\by$, $\bA$, $p_X$, and $p_{Y|Z}$, we would like to compute the MMSE estimate of $\bx$.  The GAMP algorithm \cite{Rangan_ISIT11} approaches this computationally difficult problem through a sequence of simple scalar estimation problems and matrix multiplies.
	   Remarkably, when $\bA$ is very large with IID (sub)Gaussian entries, the behavior of GAMP is rigorously characterized by a scalar state evolution \cite{Rangan_ISIT11,Javanmard_II13}.  When this state evolution has a unique fixed point, GAMP converges to the MMSE solution.  In practice, $\bA$ may not be a very large IID (sub)Gaussian matrix, in which case the theoretical guarantees of GAMP do not hold.  Still, the estimates it provides after a few (e.g., $< 25$) iterations are often very close to MMSE (see, e.g., \cite{Vila_TSP13}).
	
	   GAMP requires specification of $p_X$, which is unknown in practice.
	   To circumvent this problem, \cite{Vila_TSP13} proposed to approximate the true $p_X$ by a Gaussian-mixture with parameters $\btheta$ learned by an EM algorithm.
	   In the E-step of the EM algorithm, GAMP's posterior approximation is used in place of the true posterior, which is NP-hard to compute.
	   The resulting EM-GAMP algorithm was empirically analyzed in \cite{Vila_TSP13}, where it was shown to give similar performance to true-$p_X$ GAMP for a wide range of true $p_X$ (e.g., sparse, heavy tailed, discrete).
	   Generalizations of EM-GAMP were theoretically analyzed in \cite{Kamilov_TIT14} using the GAMP state evolution.
	
	   Algorithm~\ref{alg:EM-GAMP} details the steps of EM-GAMP.
	   Lines~1-13 are from the original GAMP algorithm and line~14 is the EM update.
	   Lines~6-7 can be interpreted as computing the posterior mean and variance of $z_i$ under the likelihood $p_{Y|Z}(y_i|z_i)$ and a pseudo-prior $z_i\sim\mathcal{CN}(\hat{p}_i,\nu_p)$, where $\hat{p}_i$ and $\nu_p$ are updated in lines~4-5.
	   Likewise, lines~12-13 can be interpreted as computing the posterior mean and variance of $x_i$ under the prior $p_X(x_i;\btheta)$ and pseudo-measurement $\hat{r}_i=x_i+\mathcal{CN}(0,\nu_r)$, where $\hat{r}_i$ and $\nu_r$ are updated in lines~8-11.
	   We note that, if $\bA$ has a fast (e.g., FFT-based) implementation, then EM-GAMP can leverage this in lines~5 and 11.
	   For more details on EM-GAMP, we refer the reader to \cite{Rangan_ISIT11,Vila_TSP13}.

	\begin{algorithm}

		\caption{The EM-GAMP algorithm} \label{alg:EM-GAMP}
		\begin{algorithmic}[1]
			\STATE \textbf{define:}
			\begin{align}
				p_{Z|Y,P}(z_i|y_i, \hat{p}_i; \nu_{p}) & \triangleq \frac{p_{Y|Z}(y_i|z_i) \mathcal{CN}(z_i; \hat{p}_i, \nu_p)}{\int_z p_{Y|Z}(y_i|z) \mathcal{CN}(z; \hat{p}_i, \nu_{p})}
                                \label{eq:pZYP}\\
				p_{X|R}(x_i|\hat{r}_i; \nu_{r},\btheta) & \triangleq \frac{p_{X}(x_i;\btheta) \mathcal{CN}(x_i; \hat{r}_i, \nu_{r})}{\int_x p_{X}(x;\btheta) \mathcal{CN}(x; \hat{r}_i, \nu_{r})}
                                \label{eq:pXR}
			\end{align}
			\STATE \textbf{initialize:}
				$\hat{\bs} = \mathbf{0}$,
                                $\btheta$,
                                $\hat{x}_i = \int_x x\, p_{X}(x;\btheta)~\forall i$,
				$\nu_x =  \int_x |x - \hat{x}_1|^2 p_{X}(x;\btheta)$,
			\FOR {$k=0,1,\dots,N_{\mr{max}}$}
			\STATE $\nu_p  \gets \|\bA\|_F^2 \nu_x/\Ny$
				
			\STATE $\hat{\bp}  \gets \bA \hat{\bx} - \nu_p\hat{\bs}$ \label{eq:multiplication}
			\STATE $\nu_{z}  \gets \Ny^{-1}\sum_{i=1}^{\Ny}\mr{Var}_{Z|Y,P} \big[z_i \,\big|\, y_i,\hat{p}_i;\nu_p \big]$ \label{eq:z_var}
			\STATE $\hat{z}_i \gets \mathbb{E}_{Z|Y,P} \big[z_i \,\big|\,y_i, \hat{p}_i;\nu_p\big], \quad \forall i \label{eq:z_hat}$
			\STATE $\nu_{s} \gets \big(1 - \nu_z/\nu_p  \big)/\nu_p$
			\STATE $\hat{s}_i \gets \nu_p^{-1} \left(\hat{z}_i - \hat{p}_i \right), \quad \forall i$
		  	\STATE $\nu_r^{-1}  \gets \|\bA\|_F^2 \nu_{s}/\Nx$
			\STATE $\hat{\br}  \gets  \hat{\bx} + \nu_r \bA^*\,\hat{\bs} \label{eq:multiplication_h}$
		  	\STATE $\nu_x \gets \Nx^{-1} \sum_{i=1}^{\Nx} \mr{Var}_{X|R} \big[x_i \,\big|\,\hat{r}_i;\nu_{r},\btheta \big]$
			\STATE $\hat{x}_i \gets \mathbb{E}_{X|R}\big[x_i \,\big|\, \hat{r}_i;\nu_{r},\btheta\big], \quad \forall i$
			\STATE update the parameters $\btheta$ using EM algorithm; 
			\ENDFOR
		        \STATE{\textbf{return} $\hat{\bx}$.}
		\end{algorithmic}
	\end{algorithm}	
	
                        \subsubsection{EM-VAMP}
                        The VAMP algorithm \cite{Rangan_arxiv16, Schniter_Asilomar16} aims to solve exactly the same problem targeted as GAMP and---like GAMP---is rigorously characterized by a scalar state-evolution.  But VAMP's state evolution holds for a much broader class of matrices $\bA$: those that are right-rotationally invariant (RRI).  For $\bA$ to be RRI, its SVD $\bA = \bU_A \bS_A \bV_A^*$ should have $\bV_A$ drawn uniformly from the set of unitary matrices; there is no restriction on $\bU_A$ and $\bS_A$.  For MMSE estimation, VAMP must be given the prior $p_X$ and likelihood $p_{Y|Z}$.  When they are not available, approximations can be learned through the EM methodology, as described in \cite{Fletcher_ICASSP17}.

                        Algorithm~\ref{alg:EM-VAMP} details the steps of EM-VAMP.  As can be seen, it alternates between nonlinear scalar estimations (lines~4-11), linear vector estimations (lines~12-19), and an EM update (line~20).
                        For scalar estimation, the conditional mean and variance of $x_i$ under prior $p_X(x_i)$ and pseudo-measurements $r_{1i}=x_i+\mathcal{CN}(0,\nu_1)$ are computed in lines~5-6, and the conditional mean and variance of $z_i$ under likelihood $p_{Y|Z}(y_i|z_i)$ and pseudo-prior $z_i\sim \mathcal{CN}(p_{1i},\tau_1)$ are computed in lines~9-10.
                        Lines~13 and 17 then compute the joint MMSE estimate of $\bx$ and $\bz$ under the pseudo-prior
	\begin{align}
		\begin{bmatrix}
			\bx \\
			\bz
		\end{bmatrix}
		&\sim \mathcal{CN} \left(
		\begin{bmatrix}
			\br_{2} \\
			\bp_{2}
		\end{bmatrix},
		\begin{bmatrix}
			\nu_{2}\bI \\
			& \tau_{2}\bI
		\end{bmatrix}
		\right)
	\end{align}
                        and the constraint $\bz=\bA\bx$.
                        We note that EM-VAMP can leverage fast implementations of $\bU_A$ and $\bV_A$ if they exist.
                        For more details, we refer the reader to \cite{Rangan_arxiv16, Schniter_Asilomar16,Fletcher_ICASSP17}.

\begin{algorithm}
	\caption{The EM-VAMP algorithm} \label{alg:EM-VAMP}
	\newcommand{\Kit}{K}
	\newcommand{\kp}{k\!+\!}
	\begin{algorithmic}[1]  \label{alg:vamp_glm}
		\STATE \textbf{define:} $p_{Z|Y,P}$ and $p_{X|R}$ from \eqref{eq:pZYP}-\eqref{eq:pXR}.
		\STATE \textbf{initialize:}
		{$\mathbf{r}_{1},\mathbf{p}_{1},\nu_{1},\tau_{1}$ and $\btheta$.}
		\FOR{$k=0,1,\dots,N_{\max}$}
		\STATE{// Scalar estimation of $x_i$}
		\STATE{$\hat{x}_{1i} \gets \mathbb{E}_{X|R}[x_i \,|\,r_{1i}; \nu_1,\btheta ],\quad \forall i$}
		\label{line:x1g}
		\STATE{$\alpha_{1} \gets \nu_1^{-1}\Nx^{-1} \sum_{i=1}^{\Nx} \mr{Var}_{X|R} \left[x_i\,|\,r_{1i}; \nu_1,\btheta \right]$}
		\STATE{$\mathbf{r}_{2} \gets (\hat{\bx}_{1} - \alpha_{1} \br_1)/(1-\alpha_{1})$,\quad
		        $\nu_{2} \gets \nu_1\alpha_1/(1-\alpha_1)$}
		\STATE{// Scalar estimation of $z_i$}
                \STATE{$\hat{z}_{1i} \gets \mathbb{E}_{Z|Y,P}[z_i \,|\,y_i,p_{1i}; \tau_1 ],\quad\forall i$}
		\label{line:z1g}
                \STATE{$\beta_1 \gets \tau_1^{-1}\Ny^{-1} \sum_{i=1}^{\Ny}\mr{Var}_{Z|Y,P} \left[z_i\,|\,y_i,p_{1i}; \tau_1 \right]$}
		\STATE{$\mathbf{p}_{2} \gets (\hat{\bz}_{1} - \beta_{1}\mathbf{p}_{1})/(1-\beta_{1})$,\quad
		        $\tau_{2} \gets \tau_{1}\beta_1/(1-\beta_1)$}
		\STATE{// LMMSE estimation of $\mathbf{x}$}
		\STATE{$\hat{\bx}_{2} \!\gets\! \mathbf{V}_A (\mathbf{S}_A^*\mathbf{S}_A\nu_2/\tau_2\!+\!\mathbf{I})^{-1} \!\big( \mathbf{S}_A^*\mathbf{U}_A^*\mathbf{p}_{2}\nu_2/\tau_2 \!+\! \mathbf{V}_A^*\mathbf{r}_{2} \big)$}
		\STATE{$\alpha_{2} \gets \Nx^{-1}\sum_{n=1}^{\Nx} \tau_2/(s_n^2\nu_2 + \tau_2)$}
		\STATE{$\mathbf{r}_1 \gets (\hat{\bx}_{2} - \alpha_{2}\mathbf{r}_{2})/(1-\alpha_{2})$,\quad
		        $\nu_1 \gets \nu_{2}\alpha_2/(1-\alpha_2)$}
		\STATE{// LMMSE estimation of $\mathbf{z}$}
		\STATE{$\hat{\bz}_{2} \gets \bA\hat{\bx}_2$}
		\STATE{$\beta_{2} \gets (1-\alpha_2)\Ny/\Nx$}
		\label{line:z2g}
		\STATE{$\mathbf{p}_1 \gets (\hat{\bz}_{2} - \beta_{2}\mathbf{p}_{2})/(1-\beta_{2})$,\quad
		        $\tau_1 \gets \tau_{2}\beta_2/(1-\beta_2)$}
		\STATE update the parameters $\btheta$ using EM algorithm; 
		\ENDFOR
		\STATE{\textbf{return} $\hat{\bx}_1$.}
	\end{algorithmic}
\end{algorithm}

\revision{\subsection{EM-AMP Algorithms for mmWave Channel Estimation with Few-bit ADCs}}
        To apply EM-GAMP and EM-VAMP to mmWave channel-estimation with few-bit ADCs, we choose the (approximating) prior family as either a Bernoulli Gaussian-mixture (GM) or Bernoulli-Gaussian (BG), with unknown parameters $\btheta$.\footnote{In this paper, we model $\bx$ using an IID sparse prior. We do not exploit possible correlation within $\bx$, which may improve estimation accuracy.}
                        That is, the coefficients $x_i$ of $\bx$ are assumed to be drawn from one of the following:
		\begin{align}
		\text{GM: }& p_{X}(x_i;\btheta) = \lambda_0 \delta(x_i) + \sum_i \lambda_i \, \mathcal{CN}(x_i; \mu_i, \phi_i)~\forall i, \label{eq:GM} \\
		\text{BG: }& p_{X}(x_i;\btheta) = \lambda_0 \delta(x_i) + (1-\lambda_0) \mathcal{CN}(x_i; 0, \phi)~\forall i \label{eq:BG} ,
		\end{align}
		        where $\lambda_0=\mathrm{Prob}\{x=0\}$ and $\left\{\lambda_i\right\}$, $\left\{\mu_i\right\}$, $\left\{\phi_i\right\}$ are the weights, means, and variances of the Gaussian mixture, respectively (which are all included in $\btheta$), and $\delta(\cdot)$ is the Dirac delta distribution.
                \revision{Since GM has more degrees of freedom than BG, it can form a better fit to the true channel distribution and thus leads to better estimation performance than BG. As shown in the simulations, the complexity of GM is a bit higher than BG since more parameters have to be estimated.}
                The expressions for $\mathbb{E}_{X|R}[x_i \,|\,\hat{r}_1; \nu_r,\btheta ]$ and $\mr{Var}_{X|R} \left[x_i\,|\,\hat{r}_i; \nu_r,\btheta \right]$ needed in lines~12-13 of Algorithm~\ref{alg:EM-GAMP} and lines~5-6 of Algorithm~\ref{alg:EM-VAMP} can be found in \cite{Vila_TSP13}, as can the EM-update expressions for $\btheta$.

		For the few-bit quantizer \eqref{eq:few-bit_ADC}, the likelihood $p_{Y|Z}$ is
                \begin{align}
		p_{Y|Z}(y_i|z_i) & \triangleq  \mathrm{Prob}\left\{y_i= \mathcal{Q}(z_i + w_i)\,|\,z_i \right\} \\
		&= \int_{w \in \mathcal{Q}^{-1}(y_i)-z_i} \hspace{-5mm} \mathcal{CN}(w;0, \sigma_w^2) .
                \end{align}
                The expressions for $\mathbb{E}_{Z|Y,P}[z_i\,|\,y_i,\hat{p}_i;\nu_p]$ and $\mr{Var}_{Z|Y,P}[z_i\,|\,y_i,\hat{p}_i;\nu_p]$ needed in lines~6-7 of Algorithm~\ref{alg:EM-GAMP} and lines~9-10 of Algorithm~\ref{alg:EM-VAMP} can be obtained by following the procedures in \cite[Chapter 3.9]{Rasmussen_Book06}.  Further details can be found in \cite[Appendix A]{Wen_Chao-Kai_TSP16}.

\bigskip
		\subsection{Computational Issues and Training Sequence Design} \label{sec:pilot_design}
                In practical mmWave applications, the dimensions $\Ny=\Nr \Np$ and $\Nx=\Nt \Nr L$ of the matrix $\bA$ are expected to be very large.
                \revision{For example, in our simulations, we consider $\Nt=\Nr=64$, $\Np=1024$, and $L=16$, which yield $\Ny=\Nx=65536$.  Storing such an $\bA$ as an explicit matrix using 4-bytes each for the real and imaginary components
would require 32 GB of memory, which is inconvenient in many applications.}
                Even when $\bA$ fits in memory, the computational complexity of EM-GAMP and EM-VAMP (or any known algorithm, for that matter) will be impractical if $\bA$ is treated as an explicit matrix, due to, e.g., per-iteration matrix-vector multiplies with $\bA$ and $\bA^*$.

                For EM-GAMP, these problems are avoided if $\bA$ and $\bA^*$ can be represented as implicit fast operators.
                Likewise, for EM-VAMP, $\bU_A$, $\bU_A^*$, $\bV_A^*$, and $\bV_A$ should be fast operators.
                Recalling from \eqref{eq:1bit_CV} that $\bA=\bC\tran \otimes \bB_{\Nr}$ with Fourier $\bB_{\Nr}$ and $\bT$-dependent $\bC\tran$, we see that the training sequence $\bT$ will determine whether $\bA$ is a fast operator.
                Thus, the design of the training signal plays a vital role in the practical implementability of mmWave channel estimation.

\black
		There are, in fact, three considerations for the design of the training signal $\bT$:
		\begin{enumerate}
			\item For GAMP to be computationally efficient, $\bA$ and $\bA^*$ should be fast operators, and for VAMP to be computationally efficient, $\bU_A$, $\bU_A^*$, $\bV_A^*$, $\bV_A$ should be.
			\item For GAMP to converge to a good solution, $\bA$ should be sufficiently dense and have sufficiently low peak-to-average squared-singular-value ratio \cite{Rangan_ISIT14}.
			\item To improve the efficiency of the power amplifier at the transmitter, the elements of $\bT$ should have low \ac{PAPR}.
		\end{enumerate}

                \subsubsection{Training Structure}
                To satisfy these three considerations, we propose to structure $\bT\in\Complex^{\Nt\times\Np}$ as follows.
                Denote the first row of $\bT$, i.e., the signal sent by the first transmit antenna, as $\bt\tran \triangleq \left[t\left[0\right],\dots, t\left[\Np-1 \right] \right]$.
                We fix $\Np$ at an integer multiple of $\Nt L$ and construct $\bT$ such that its $n$th row (for $n\geq 0$) is the $nL$-place circular shift of $\bt\tran$.  That is, each antenna sends a circularly shifted version of the signal transmitted by the first antenna.
                The elements of $\bT$ (for $m\geq 0$) are then
	\begin{align} \label{eq:T_n_m}
	[\bT]_{n,m} & = t[\langle m - nL \rangle_{\Np}],
	\end{align}
	        where $\langle \cdot \rangle_{\Np}$ is the modulo-$\Np$ remainder.
	        With the construction in \eqref{eq:T_n_m}, $\bT \bJ_0$ from \eqref{eq:tilde_T}
                contains the $\left\{0, L, 2L,..., (\Nt-1)L\right\}$ shifts of $\bt$.
                Likewise, $\bT \bJ_1$ contains the $\left\{1, L+1, 2L+1,..., (\Nt-1)L+1\right\}$ shifts of $\bt$, and $\bT \bJ_{L-1}$ contains the $\left\{L-1, 2L-1, 3L-1,..., \Nt L-1\right\}$ right shifts of $\bt$.
	        Altogether, the rows of $\tilde{\bT}$ in \eqref{eq:tilde_T} will consist of the first $\Nt L$ circular shifts of the sequence $\bt$. We can thus re-order the rows in $\tilde{\bT}$ to make it a Toeplitz matrix, which we will denote by $\overline{\bT}$ in the sequel.
	
                \subsubsection{Fast Implementation of $\bA\hat{\bx}$ and $\bA^*\hat{\bs}$}
        Using the structure above, $\bA\hat{\bx}$ and $\bA^*\hat{\bs}$ can be efficiently computed.
        Notice
	\begin{align}
	\lefteqn{ \mr{unvec}\left(\bA \hat{\bx} \right) }\nonumber \\
	&= \bB_{\Nr} \hat{\bX} \left( \bI_L \otimes \bB^*_{\Nt} \right) \tilde{\bT} \\
	&\stackrel{(a)}{=} \bB_{\Nr} \hat{\bX} \left( \bI_L \otimes \bB^*_{\Nt} \right) \bK^{(L, \Nt)} \overline{\bT} \\
	&\stackrel{(b)}{=} \bB_{\Nr} \hat{\bX} \bK^{(L, \Nt)} \bK^{(\Nt, L)} \left( \bI_L \otimes \bB^*_{\Nt} \right) \bK^{(L, \Nt)} \overline{\bT} \\
	&\stackrel{(c)}{=} \bB_{\Nr} \hat{\bX} \bK^{(L, \Nt)} \left(\bB^*_{\Nt} \otimes \bI_L \right) \overline{\bT} \\
	&\stackrel{(d)}{=} \bB_{\Nr} \left[ \overline{\bT}\tran \left(\bB^*_{\Nt} \otimes \bI_L \right) \overline{\bX} \right]\tran , \label{eq:Ax_fast}
	\end{align}
	where (a) follows with commutation matrix\footnote{The commutation matrix matrix $\bK^{(m,n)}$ is the $mn \times mn$ matrix which, for any $m \times n$ matrix $\bM$, transforms $\mr{vec}(\bM)$ into $\mr{vec}(\bM\tran)$, i.e., $\bK^{(m,n)} \mr{vec}(\bM) = \mr{vec}(\bM\tran)$.}
        $\bK^{(L, \Nt)} \in \mathbb{R}^{\Nt L \times \Nt L}$;
        (b) follows from $\bK^{(L, \Nt)} \bK^{(\Nt, L)}=\bI_{\Nt L}$; (c) follows from $\bK^{(\Nt, L)}\left( \bI_L \otimes \bB^*_{\Nt} \right) \bK^{(L, \Nt)} = \bB^*_{\Nt} \otimes \bI_L$; and (d) follows from $\overline{\bX} \triangleq (\hat{\bX} \bK^{(L, \Nt)})\tran$, $(\bA \otimes \bB)\tran=\bA\tran \otimes \bB\tran$, and the symmetry of $\bB^*_{\Nt}$.
	Note that $\hat{\bX} \bK^{(L, \Nt)}$ is merely a reordering of the columns in $\hat{\bX}$.

        We now show that \eqref{eq:Ax_fast} has a fast implementation.
        First, for any $\bv \in \mathbb{C}^{L\Nt \times 1} $, notice that
	\begin{align} \label{eq:B_I_v}
	\left(\bB^*_{\Nt} \otimes \bI_L \right) \bv = \mr{vec} \left(\bV \bB^*_{\Nt}\right) = \mr{vec} \left( \left( \bB_{\Nt} \bV^* \right)^*\right),
	\end{align}
	where $\bV \in \mathbb{C}^{L \times \Nt}$ is the column-wise matricization of $\bv$. Second, for any $\bu \in \mathbb{C}^{\Nt \times 1}$, notice that
	\begin{align} \label{eq:B_u}
	\bB_{\Nt} \bu = \left( \bF_{\Nta}  \otimes \bF_{\Nte} \right) \bu = \mr{vec} \left( \bF_{\Nte} \bU \bF_{\Nta} \right),
	\end{align}
	where $\bU \in \mathbb{C}^{\Nte \times \Nta}$ is the column-wise matricization of $\bu$.
        Therefore, each of the $\Nr$ columns of the multiplication $\left(\bB^*_{\Nt} \otimes \bI_L \right) \overline{\bX}$ in \eqref{eq:Ax_fast} can be accomplished by $\Nte$-point and $\Nta$-point FFTs.\footnote{To see this, note that each column of the multiplication $(\bB^*_{\Nt} \otimes \bI_L ) \overline{\bX}$ can be represented as \eqref{eq:B_I_v} with appropriate $\bv$, and thus computed via $\bB_{\Nt} \bV^*$ due to \eqref{eq:B_I_v}.  Then, $\bB_{\Nt} \bV^*$ can be computed columnwise through subproblems of the form \eqref{eq:B_u} with appropriate $\bu$, which in turn can be tackled through $\Nte$-point FFTs of the columns of $\bU$ and $\Nta$-point FFTs of the rows of $\bU=\mr{unvec}(\bu)$, as shown by the right side of \eqref{eq:B_u}.}
	The result of that multiplication is then left-multiplied by $\overline{\bT}\tran$, which can be performed via fast convolution using an $\Np$-point FFT (since $\overline{\bT}\tran$ contains the first $\Nt L$ columns of an $\Np\times\Np$ circulant matrix). Finally, the left-multiplication by $\bB_{\Nr}$ can be performed, as in \eqref{eq:B_u}, using $\Nra$-point and $\Nre$-point FFTs. In summary, there are a total of $\Nr L$ FFTs of length $\Np$, one inverse-FFT of length $\Np$, $\Nta \Nr$ FFTs of length $\Nte$, $\Nte \Nr$ FFTs of length $\Nta$, $\Nra \Np$ FFTs of length $\Nre$, and $\Nre \Np$ FFTs of length $\Nra$.
        Multiplications of the form $\bA^*\hat{\bs}$ can be computed similarly.
        Note also that the memory footprint of $\bA$ reduces to that of $\bt\in\Complex^{\Np}$.

                \subsubsection{Choice of the training sequence $\bt$}
	For GAMP to work well, we want the measurement matrix $\bA$ to have sufficiently low peak-to-average squared-singular-value ratio \cite{Rangan_ISIT14}.
        Likewise, for VAMP to be fast, we want the singular-vector matrices $\bU_A$ and $\bV_A$ of $\bA$ to be fast operators.
        We now show that both concerns can be addressed through the design of $\bt$.

	From \eqref{eq:tilde_T}, \eqref{eq:output_z}, and \eqref{eq:1bit_CV}, the matrix $\bA$ can be written as
	\begin{align}
		\bA &= \left( \left( \bI_L \otimes \bB^*_{\Nt} \right) \tilde{\bT} \right)\tran \otimes \bB_{\Nr}.
	\end{align}
	Since $\bI_L \otimes \bB^*_{\Nt}$ and $\bB_{\Nr}$ are unitary matrices, the singular values of $\bA$ have the form
	\begin{align}
		\underbrace{\lambda_1, \dots, \lambda_1}_{\Nr}, \underbrace{\lambda_2, \dots, \lambda_2}_{\Nr}, \dots, \underbrace{\lambda_{\Nt L}, \dots, \lambda_{\Nt L}}_{\Nr},
	\end{align}
	where $\lambda_1, \lambda_2, \dots, \lambda_{\Nt L}$ are the singular values of $\tilde{\bT}$. Therefore, we want $\tilde{\bT}$ to have sufficiently low peak-to-average squared-singular-value ratio.
	One way to ensure this property is to choose $\bt$ as a Zadoff-Chu (ZC) sequence \cite{Chu_IT72}.

        A length-$\Np$ ZC sequence is defined as
	\begin{align}
	t[k] =
	\begin{cases}
	\sqrt{\frac{\Pt}{\Nt}} \exp \left( \j \frac{\pi k(k+1)}{\Np} \right) & \text{if $\Np$ is odd,} \\
	\sqrt{\frac{\Pt}{\Nt}} \exp \left( \j \frac{\pi k^2}{\Np} \right) & \text{if $\Np$ is even.}
	\end{cases}
	\end{align}
	The periodic autocorrelation of a ZC sequence equals a scaled Kronecker delta. As a result, $\tilde{\bT} \tilde{\bT}^* = \frac{\Pt \Np}{\Nt} \bI_{\Nt L}$, and so $\lambda_1=\lambda_2= \cdots=\lambda_{\Nt L}=\sqrt{\Pt \Np/\Nt}$.
	Consequently, $\bA$ will have the minimum possible peak-to-average squared-singular-value ratio, which is good for GAMP.
    Furthermore, this singular-value structure implies that the singular-vector matrices can be chosen as $\bV_A=\bI$ and $\bU_A = \sqrt{\frac{\Nt}{\Pt\Np}} \bA$. Since $\bV_A$ and $\bU_A$ have fast implementations, they are good for VAMP. 
	

	\revision{An additional benefit of choosing the ZC sequence for $\bt$ is its constant-modulus property, which ensures that $\bT$ has a low peak-to-average power ratio \ac{PAPR}. For this reason, ZC sequences are currently used as reference signals in LTE\cite{TS36_211}.}
        \subsubsection{Other choices of $\bT$}
        In Section \ref{sec:simulation}, we investigate other training designs, for example, $\bT$ composed of Golay complementary sequences, IID QPSK entries, and IID Gaussian entries.  Although their recovery performance is comparable to our proposed design, they lead to GAMP and VAMP algorithms with much higher complexity due to the lack of fast methods to compute the matrix-vector multiplications with $\bA,\bA^*,\bU_A,\bU_A^*,\bV_A,\bV_A^*$.
\black
	
	\subsection{Benchmark Algorithms} \label{sec:baseline}
		In this section, we describe several other signal reconstruction approaches that we will use as benchmarks.
		The first approach is the \ac{LS} estimator adopted in \cite{Wang_Shengchu_TWC15, Risi_arxiv14, Jacobsson_ICC15}. In this case
		\begin{align}
		\hat{\bX}_{\mr{LS}}
                = \bB^*_{\Nr} \bY \bC^{\dagger},
		\end{align}
		where $\bC^{\dagger}$ is the Moore-Penrose pseudo-inverse of $\bC$.
                When $\bC$ has full row-rank, we have
 		\begin{align}
 		\hat{\bX}_{\mr{LS}} = \bB^*_{\Nr} \bY \bC^* \left(\bC \bC^*\right)^{-1}.
                \label{eq:LS}
 		\end{align}

                Another approach is obtained by linearizing the quantizer and applying LMMSE \cite{Li_Yongzhi_TSP17,Mollen_TWC17}.
                We explain the linearization procedure because it will be used later for achievable-rate analysis.
		Using Bussgang's theorem, the quantizer output $\by$ can be decomposed into a signal component plus a distortion $\bwq$ that is uncorrelated with the signal component \cite{Mezghani_ISIT12}, i.e.,
		\begin{align}
		\by &= \mathcal{Q} \left( \bA \bx  + \bw \right) \nonumber \\
		&= (1 - \eta_b) \left( \bA \bx + \bw \right) + \bwq, \label{eq:AQNM}
		\end{align}
		where $\eta_b$ is the NMSE given in Table~\ref{tab:Delta}.
                Under additional mild assumptions, \cite[eq.(30)]{Mezghani_ISIT12} showed that
                the ``effective'' noise $\hat{\bw} \triangleq (1-\eta_b)\bw + \bwq$ in the linearized model
		\begin{align}
		\by = (1 - \eta_b)  \bA \bx + \hat{\bw}
                \label{eq:linearized}
		\end{align}
                has covariance $\bSigma_{\hat{w}}$ that is well approximated as
		\begin{align}
                \bSigma_{\hat{w}}
                &\approx (1-\eta_b)\Big( (1-\eta_b)\mathbb{E}\left[ \bw \bw^* \right]
                        \nonumber\\&\quad
                        + \eta_b\mr{diag}\big(\mathbb{E}\left[\bA \bx \bx^* \bA^* \right]
                                               +\mathbb{E}\left[ \bw \bw^* \right]\big)
                        \Big) \triangleq \hat{\bSigma}_{\hat{w}},
                \label{eq:white}
		\end{align}
		where $\mathrm{diag} (\bM)$ is formed by zeroing the off-diagonal elements of $\bM$.
                If we furthermore assume $\mathbb{E}[\bx \bx^*] = \sigma_x^2 \bI $ with $\sigma_x^2$ given in \eqref{eq:sigmaxhat} then, due to the independence of $\bx$ and $\bA$,
		\begin{align}
                \lefteqn{ \hat{\bSigma}_{\hat{w}}
                = (1-\eta_b) \sigma_w^2 \bI_{\Nr\Np}+ \eta_b(1-\eta_b)\sigma_x^2\mr{diag}\big(\mathbb{E}[\bA \bA^*] \big) }\\
                &\stackrel{(a)}{=} (1-\eta_b) \sigma_w^2 \bI_{\Nr\Np} + \eta_b(1-\eta_b)\sigma_x^2\mr{diag}\big(\mathbb{E}[\tilde{\bT}^* \tilde{\bT}]\tran \otimes \bI_{\Nr}\big) \nonumber \\
		&= \big( (1-\eta_b) \sigma_w^2 + \eta_b(1-\eta_b) \Pt L \sigma_x^2\big) \bI_{\Nr \Np},
                \label{eq:Sighatw}
		\end{align}
                where (a) uses
                $\bA \bA^*
                = (\bC^*\bC)\tran \otimes \bI_{\Nr}
                = (\tilde{\bT}^*\tilde{\bT})\tran \otimes \bI_{\Nr}$, which follows from
                from \eqref{eq:1bit_CV}, the unitary property of $\bB_{\Nr}$,
                and \eqref{eq:tilde_T}-\eqref{eq:output_z},
                and \eqref{eq:Sighatw} uses
                $\mathrm{diag}(\mathbb{E}[\tilde{\bT}^*\tilde{\bT}])=\Pt L\bI_{\Np}$,
                which follows from the transmitter power constraint.
                Thus, the effective noise $\hat{\bw}$ can be approximated as spectrally white with variance
		\begin{align}
		\sigma_{\hat{w}}^2
		&= (1-\eta_b)\big( \sigma_w^2 + \eta_b \Pt L \sigma_x^2\big). \label{eq:effective_noise_var}
		\end{align}
                The approximation \eqref{eq:white} has been shown to be quite accurate for MIMO communication, especially at low SNR \cite{Mezghani_ISIT12, Roth_arxiv16}.
		Note, however, that $\hat{\bw}$ is non-Gaussian.

                Assuming $\mathbb{E}[\bx \bx^*] = \sigma_x^2 \bI$ (as above) and leveraging the result \eqref{eq:Sighatw} that
                $\mathbb{E}[\hat{\bw}\hat{\bw}^*] \approx \sigma_{\hat{w}}^2\bI$, one can straightforwardly derive the \ac{LMMSE} estimator of $\bx$ from $\by$ in \eqref{eq:linearized}.  We will refer to it as the ``\ac{ALMMSE} estimator'' due to the approximation \eqref{eq:white}.  Expressed in terms of $\bX=\mr{unvec}(\bx)$, it takes the form
		\begin{align}
		\lefteqn{ \hat{\bX}_{\mr{ALMMSE}} }\\
		&= \bB_{\Nr}^* \bY  \bC^* \left( \left(1- \eta_b \right) \bC \bC^*
                        + \left( \frac{\sigma_w^2}{\sigma_x^2} + \eta_b \Pt L　\right)
                        \bI_{\Nt L} \right)^{-1},  \nonumber
		\end{align}
                which is similar to \eqref{eq:LS} but with a regularized inverse.
        \black	
		
	A third benchmark algorithm follows by applying sparse reconstruction to the linearized model \eqref{eq:linearized}, which---unlike the methods above---leverages the fact that $\bx$ is approximately sparse due to leakage. In particular, we used the ``SPGL1'' algorithm \cite{Berg_SIAM08} to
        solve the basis pursuit denoising (BPDN) problem
	\begin{align}
        \arg\min \|\bx \|_1
	~\mathrm{s.t.}~ \| \by - (1 - \eta_b) \bA \bx \|^2 \leq  \sigma_{\hat{w}}^2 \Np \Nr,
	\end{align}
	with $\sigma_{\hat{w}}^2$ defined in \eqref{eq:effective_noise_var}.\footnote{We also tried the $\ell_1$-based method from \cite{Zymnis_SPL10}, but we experienced numerical problems with their first-order method.  Since our $\bA$ was too large to fit in memory, we could not use the CVX implementation of \cite{Zymnis_SPL10} nor \cite{Plan_IT13}.}
        Similar to EM-GAMP and EM-VAMP, the computational complexity of SPGL1 is dominated by matrix-vector multiplications with $\bA$ and $\bA^*$, and so the fast implementation \eqref{eq:Ax_fast} is used.
	
	A fourth benchmark algorithm is the quantized iterative hard thresholding (QIHT) algorithm proposed in \cite{Jacques_IT13,Jacques_arxiv13}. At iteration $k=1,2,\dots$, the estimate is updated as
	\begin{align}
	\hat{\bs}^{(k)}	&= \mathcal{Q}\left(\by - \bA \hat{\bx}^{(k)}\right),\\
	\hat{\bx}^{(k+1)} &= h_K\left(\hat{\bx}^{(k)} + \tau \bA^* \hat{\bs}^{(k)} \right),
	\end{align}
	where $h_K(\bx)$ is the hard thresholding operator that zeros all but the $K$ largest (in magnitude) components of $\bx$. For convergence, it is suggested that $\tau < \|\bA\|_2^{-2}$ for spectral norm $\|\bA\|_2$. Our simulations used
        $\hat{\bx}^{(1)}=\mathbf{0}$, $\tau = 0.1 \frac{\Nt}{\Pt \Np}$, and
        $K=\Nx /100$.  That is, QIHT kept only the largest $1\%$ of the elements in each iteration.

	\subsection{Norm Estimation of the Channel}
	With a one-bit ADC, all amplitude information is lost during quantization. Therefore, it is difficult to precisely recover the channel norm $\|\bx\|$, especially at high SNR. In previous work \cite{Choi_TCOM16,Wen_Chao-Kai_TSP16}, the variance of the channel was assumed to be known to avoid this issue. However, it is possible in practice to estimate the channel norm from the average signal power received by the antenna circuit before quantization, for example, the AGC circuit.
	
	Let us use $\hat{\bz} \triangleq ( \bC\tran \otimes \bB_{\Nr} ) \bx  + \bw$ to denote the noisy unquantized signal.
        We assume that it is possible to measure the total received power across all antennas, which is $\mathbb{E}[\|\hat{\bz}\|^2]$.
        Under the assumption that $\mathbb{E}[\bx \bx^*] = \sigma_x^2 \bI $, it is straightforward to show that
        \begin{align}
        \mathbb{E}[\|\hat{\bz}\|^2]
        &= \|\bC\tran \otimes \bB_{\Nr}\|_F^2 \sigma_x^2 + \Np \Nr \sigma_{w}^2.
        \end{align}
        In the UPA case, we have $\|\bC\tran \otimes \bB_{\Nr}\|_F^2 = \Nr \|\bC\|_F^2$,
        and for ZC training we have $\|\bC\|_F^2 = \Pt\Np L$, which implies that
	\begin{align}
        \sigma_x^2
	&= \frac{ \mathbb{E}[\|\hat{\bz}\|^2] - \Np \Nr \sigma_{w}^2 }{\Pt L \Np \Nr} .
        \label{eq:sigmaxhat}
	\end{align}%
        Under $\mathbb{E}[\bx \bx^*] = \sigma_x^2 \bI $, the law of large numbers then says that
        $\|\bx\| \approx \sigma_x\sqrt{\Nr\Nt L}$ for large $\Nr\Nt L$.
	Thus, for all channel estimators, we normalized the channel estimate $\hat{\bx}$ as follows,
	\begin{align}
	\tilde{\bx} = \hat{\bx} \frac{\hat{\|\bx\|}}{ \|\hat{\bx}\|}
        \text{~~for~~}
        \hat{\|\bx\|} \triangleq \sigma_x\sqrt{\Nr\Nt L}
        , \label{eq:debiase}
	\end{align}
	since it reduced the channel estimation error in all cases.

	\section{Simulation Results} \label{sec:simulation}
\begin{figure*}[t]
			\centering
                        \newcommand{\sz}{.64}
			\subfigure[]{
			\includegraphics[width=\sz\columnwidth,clip]{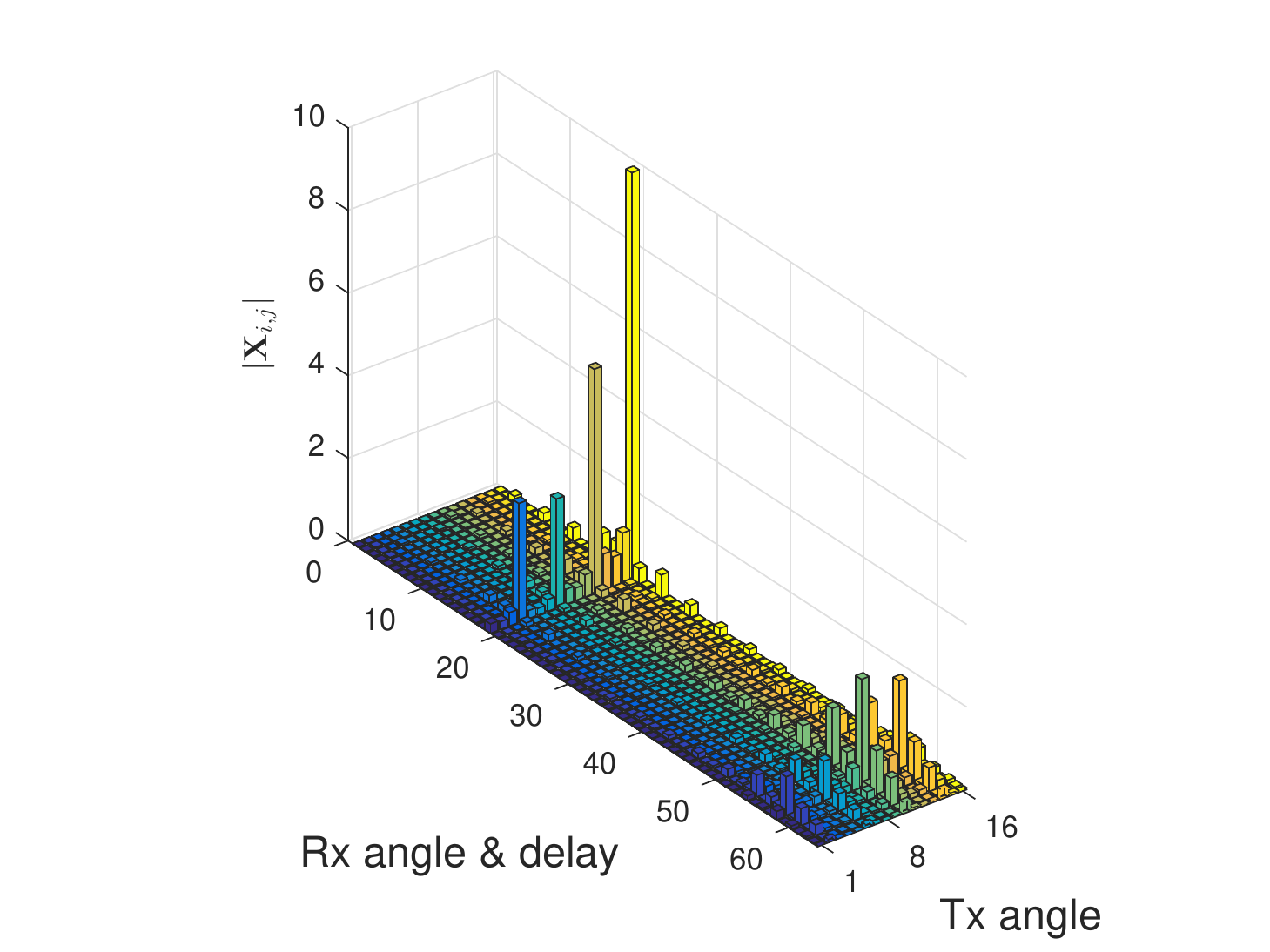}
				\label{fig:True_Channel}
				}
			\subfigure[]{
			\includegraphics[width=\sz\columnwidth,clip]{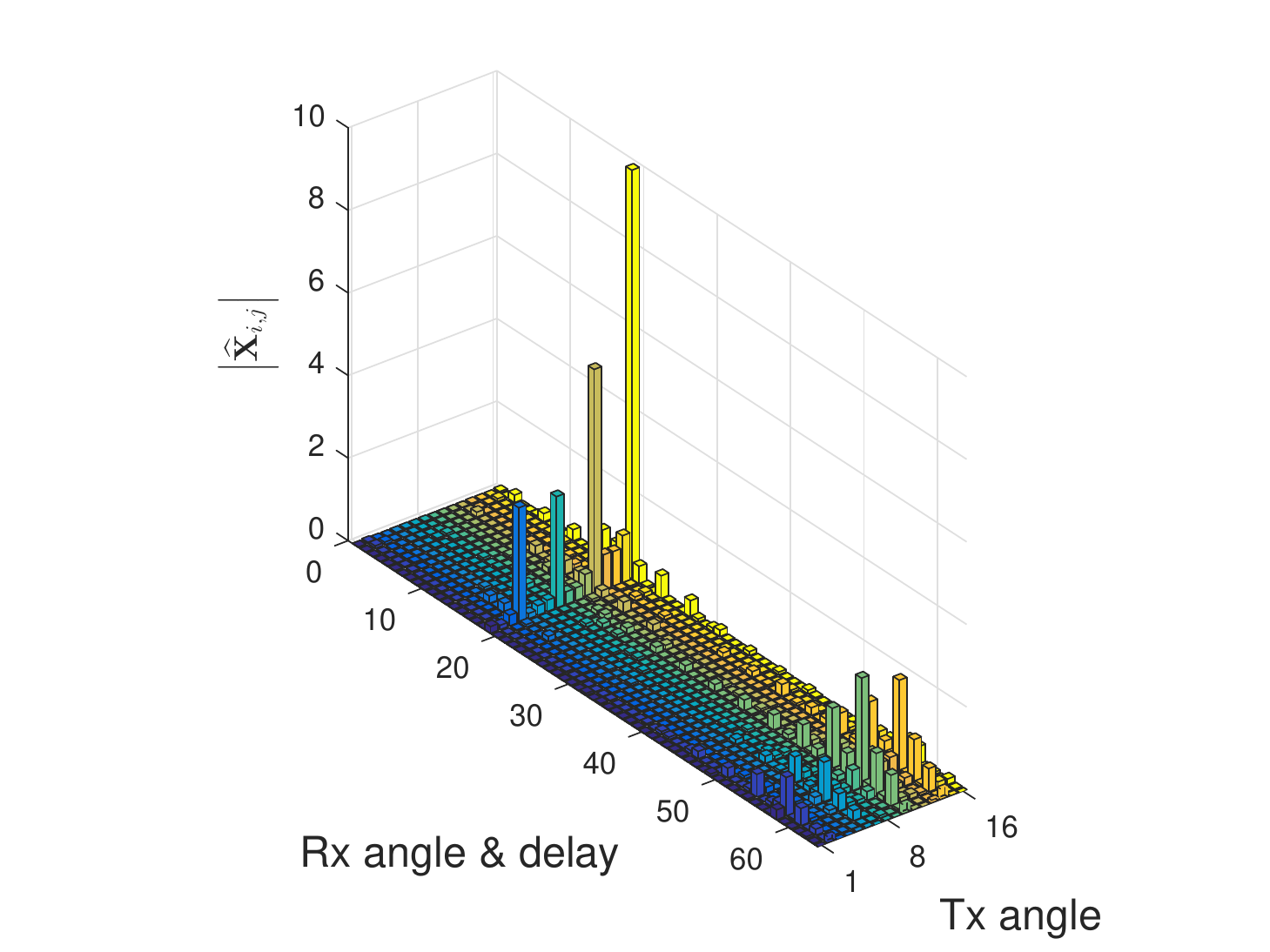}
				\label{fig:Estimated_Channel}
				}
			\subfigure[]{
			\includegraphics[width=\sz\columnwidth,clip]{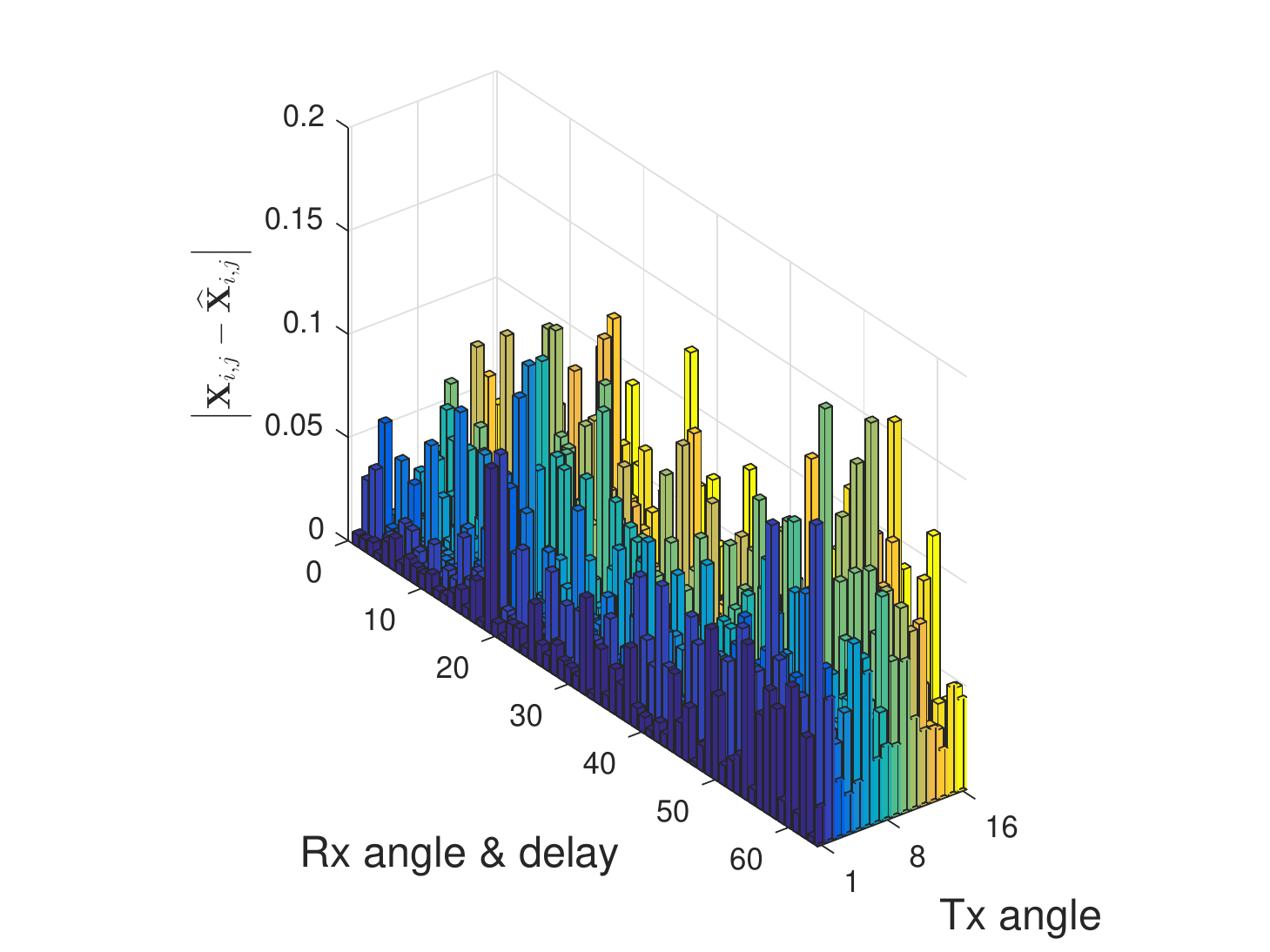}
				\label{fig:Estimation_Error}
			}
			\caption{The true angle-delay channel magnitudes are plotted in (a), the magnitudes of the EM-GM-VAMP estimate in (b), and the estimation error magnitudes in (c) \revision{in linear scale}. In particular, (a) shows $|[\bX[\ell]]_{i,j}|$ at the $(x,y)$ location $x=4\ell+i$ and $y=j$. For this example, there were $2$ clusters in a $4 \times 16$ MIMO channel with a delay spread of $16$ samples. The block length of the training signal was $512$ and the SNR was $10$~dB. The ADC had 4-bit resolution.}\label{fig:Example}
		\end{figure*}

	We first provide an illustrative example to visualize the behavior of the proposed channel estimation procedure. In Fig. \ref{fig:Example}, we show the result of estimating a broadband $\Nr \times \Nt = 4 \times 16$ MIMO channel.
	The transmitter was equipped with a $4 \times 4$ UPA and the receiver with a $2 \times 2$ UPA. The channel had $\Ncl=2$ clusters and delay spread of at most $L=16$ samples. 4-bit ADCs and raised-cosine pulse-shaping with roll-off factor zero was used.  The training length was shifted-ZC of length $\Np=512$. The true magnitudes of the angle-delay channel are plotted in \figref{fig:True_Channel} while the estimated magnitudes are plotted in Fig. 3(b).
	As seen in the figure, the EM-GM-VAMP algorithm can estimate the channel $\bX$ quite accurately: the two clusters can be easily identified in \figref{fig:Estimated_Channel} and the estimation error in \figref{fig:Estimation_Error} is small. The joint angle-delay sparsity of the mmWave channel is visible in \figref{fig:True_Channel}, as is the \emph{approximate} nature of the sparsity.

	In the following subsections, we consider a more practical scenario, where the transmitter and receiver are both equipped with $8 \times 8$ UPAs.
	We consider a channel with $4$ clusters, each of $10$ paths, with azimuth and elevation angular spread of $7.5$ degrees. The delay spread was at most $16$ samples.
	These numbers were chosen according to the urban macro (UMa) NLOS channel measurement results at $28$~GHz, as given in the white paper \cite{WP_5G_Channel_Model}.
        The results we report are the average of $100$ different channel realizations.
        The channel was normalized so that $\mathbb{E}\left[\sum_{\ell} \|\bH[\ell]\|_F^2 \right] = \mathbb{E} \left[ \| \bx \|^2 \right] = \Nt \Nr$.
        We define
        \begin{align*}
        \text{SNR}
        \triangleq \frac{\mathbb{E}[\|\bz\|^2]}{\mathbb{E}[\|\bw \|^2]}
        &= \frac{\mathbb{E}[\mathrm{tr}\left\{\bA \bx \bx^* \bA^* \right\}]}{\Nr \Np \sigma_w^2} = \frac{\Pt L \Np \Nr}{L \Nr \Np \sigma_w^2} = \frac{\Pt}{\sigma_w^2}.
        \end{align*}
	
	\subsection{Choice of Training Matrix}
\begin{figure}[t]
			\begin{centering}
				\includegraphics[width=0.95\columnwidth]{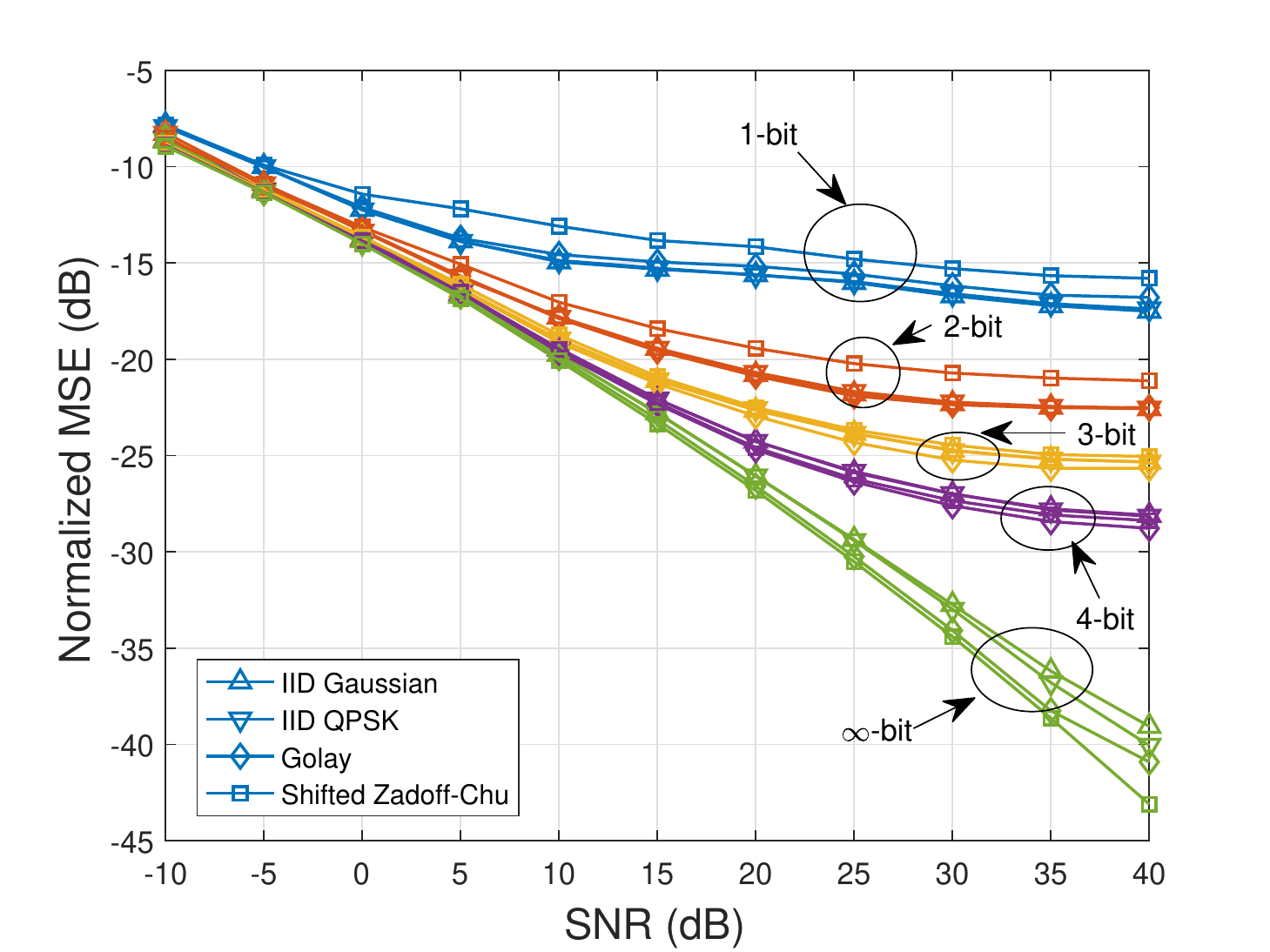}
				\vspace{-0.1cm}
				\centering
				\caption{EM-GM-VAMP recovery performance versus SNR for training matrix $\bT$ constructed from IID Gaussian entries, IID QPSK entries, Golay complementary sequences, and shifted-ZC sequences.  Here, $\Nt=\Nr=64$, $L=16$, $\Ncl=4$, and $\Np = 2048$.}\label{fig:Pilot_Comparison}
			\end{centering}
			\vspace{-0.3cm}
		\end{figure}

	We first investigate the performance of various choices of training matrix $\bT$.
        In addition to the circularly-shifted-ZC design proposed in \secref{sec:pilot_design}, we try $\bT$ constructed from Golay complementary sequences, IID random Gaussian entries, and IID random QPSK entries. The channel estimation NMSEs of EM-GM-VAMP with these $\bT$ are compared in \figref{fig:Pilot_Comparison}, where
	\begin{align}
	\text{NMSE}(\tilde{\bx}) \triangleq \mathbb{E} \left[ \frac{\|\tilde{\bx} -\bx\|^2 }{\|\bx\|^2} \right],
	\end{align}
	for $\tilde{\bx}$ normalized according to \eqref{eq:debiase}.
	As seen in the figure, the NMSEs of the four training designs are very close. However, the Golay sequence, random QPSK, and ZC sequences are preferred because of their constant modulus property, which leads to low PAPR.  Furthermore, the ZC-based design allows efficient implementation of the matrix-vector multiplications in VAMP, as discussed in \secref{sec:pilot_design}.
	
\begin{figure}[t]
			\begin{centering}
				\includegraphics[width=0.95\columnwidth]{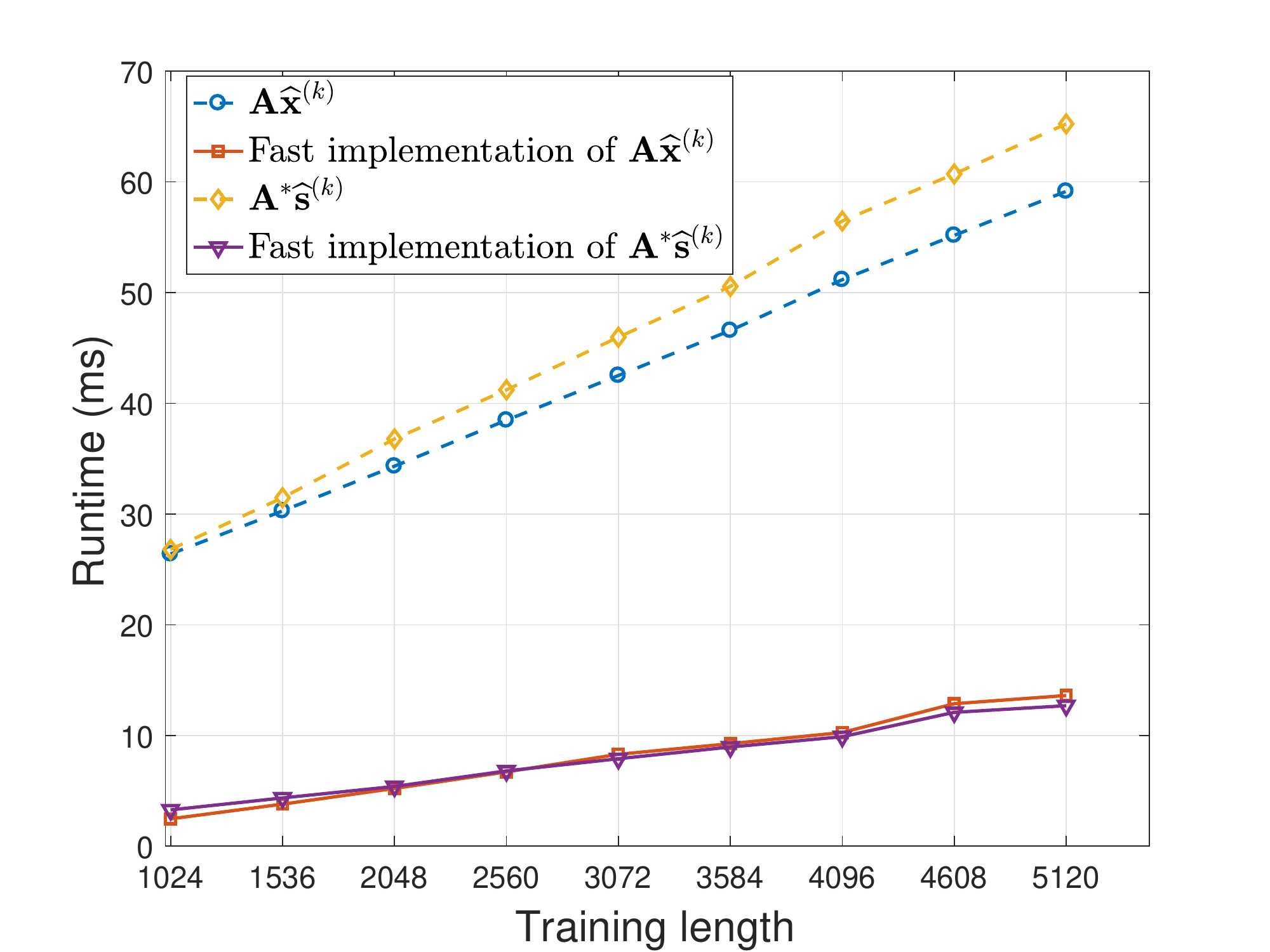}
				\vspace{-0.1cm}
				\centering
				\caption{Runtime versus training length $\Np$ to compute the matrix-vector multiplications $\bA\hat{\bx}$ and $\bA^*\hat{\bs}$ implemented via $\bA\hat{\bx}=\mr{vec} \left( \bB_{\Nr}\hat{\bX}\bC \right)$ or via the fast method from \secref{sec:pilot_design}. Here, $\Nt=\Nr=64$ and $L=16$, so that the dimensions of $\bA$ are $\left(64 \Np \right) \times 65536$. }\label{fig:Runtime_Ax}
			\end{centering}
			\vspace{-0.3cm}
		\end{figure}

	The runtime of the matrix-vector multiplications $\bA\hat{\bx}$ and $\bA^*\hat{\bs}$ and their fast implementations (using MATLAB R2016a on a standard desktop computer) are shown in \figref{fig:Runtime_Ax}.
	In the baseline implementation, $\bA\hat{\bx}$ was computed via $\mr{vec} ( \bB_{\Nr}\hat{\bX}\bC )$ to avoid generating and storing the high-dimensional matrix $\bA$, and the computation of $\bA^*\hat{\bs}$ was done similarly.
	The fast implementation was described in \secref{sec:pilot_design}.
	As seen in the figure, the fast implementation consumes much less time than the baseline implementation.
        Much faster runtimes would be possible if the FFTs were implemented in hardware.

	\subsection{Algorithm Complexity}
\begin{figure}[t]
			\begin{centering}
				\includegraphics[width=0.95\columnwidth]{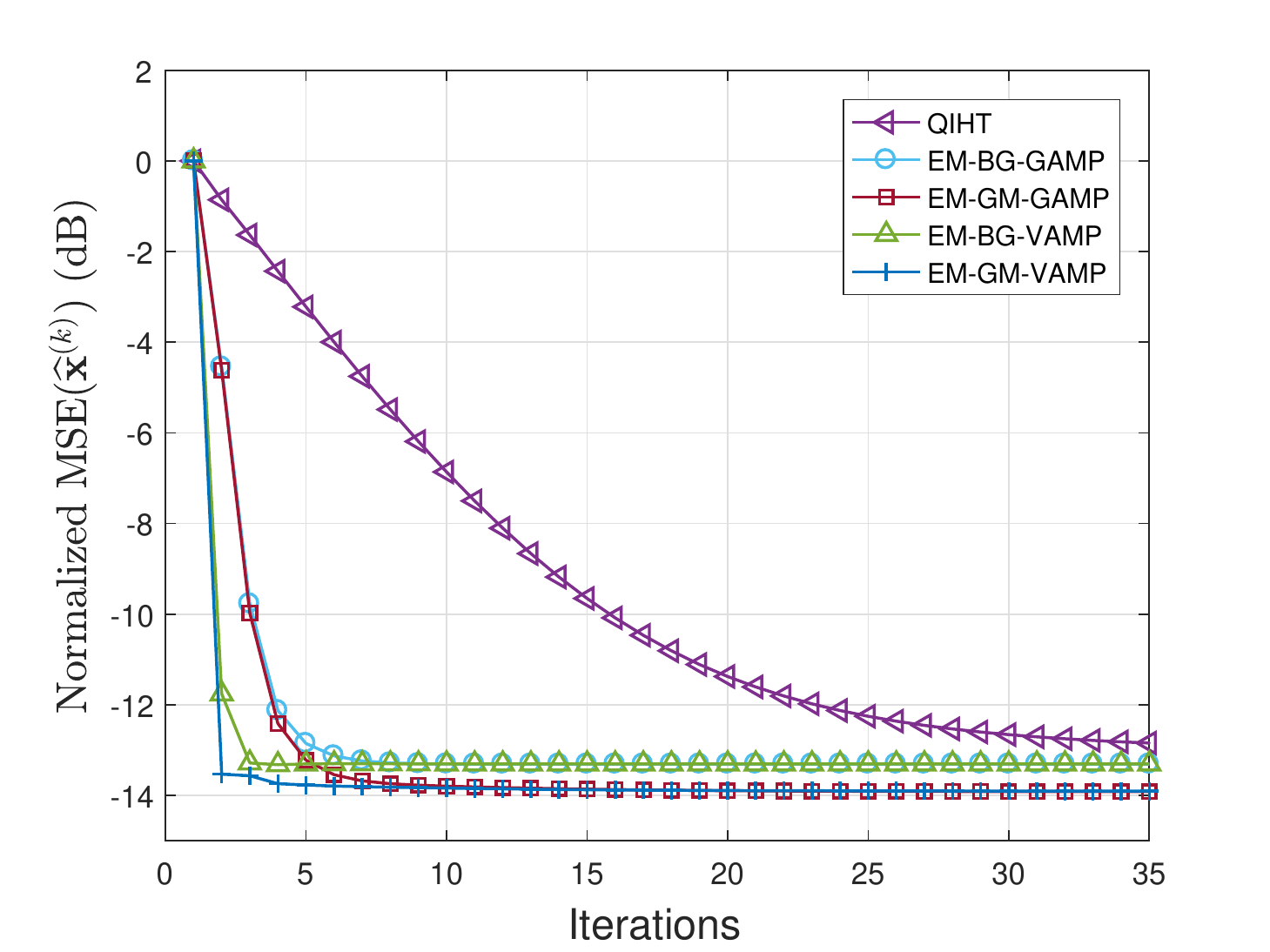}
				\vspace{-0.1cm}
				\centering
				\caption{NMSE versus iteration for several algorithms. Here, SNR $= 0$~dB, the ADC had 4 bits, and the training was shifted ZC of length $\Np=2048$.}\label{fig:NMSE_vs_Iterations}
			\end{centering}
			\vspace{-0.3cm}
		\end{figure}

        We now investigate the computational complexity of the various algorithms under test.
        \figref{fig:NMSE_vs_Iterations} shows NMSE versus iteration for EM-GAMP, EM-VAMP, and QIHT.
	The figure shows the VAMP algorithms converging in $\approx 3$ iterations, the GAMP algorithms converging in $\approx 8$ iterations, and QIHT converging in $\approx 35$ iterations.
	
\begin{figure}[t]
			\begin{centering}
				\includegraphics[width=0.95\columnwidth]{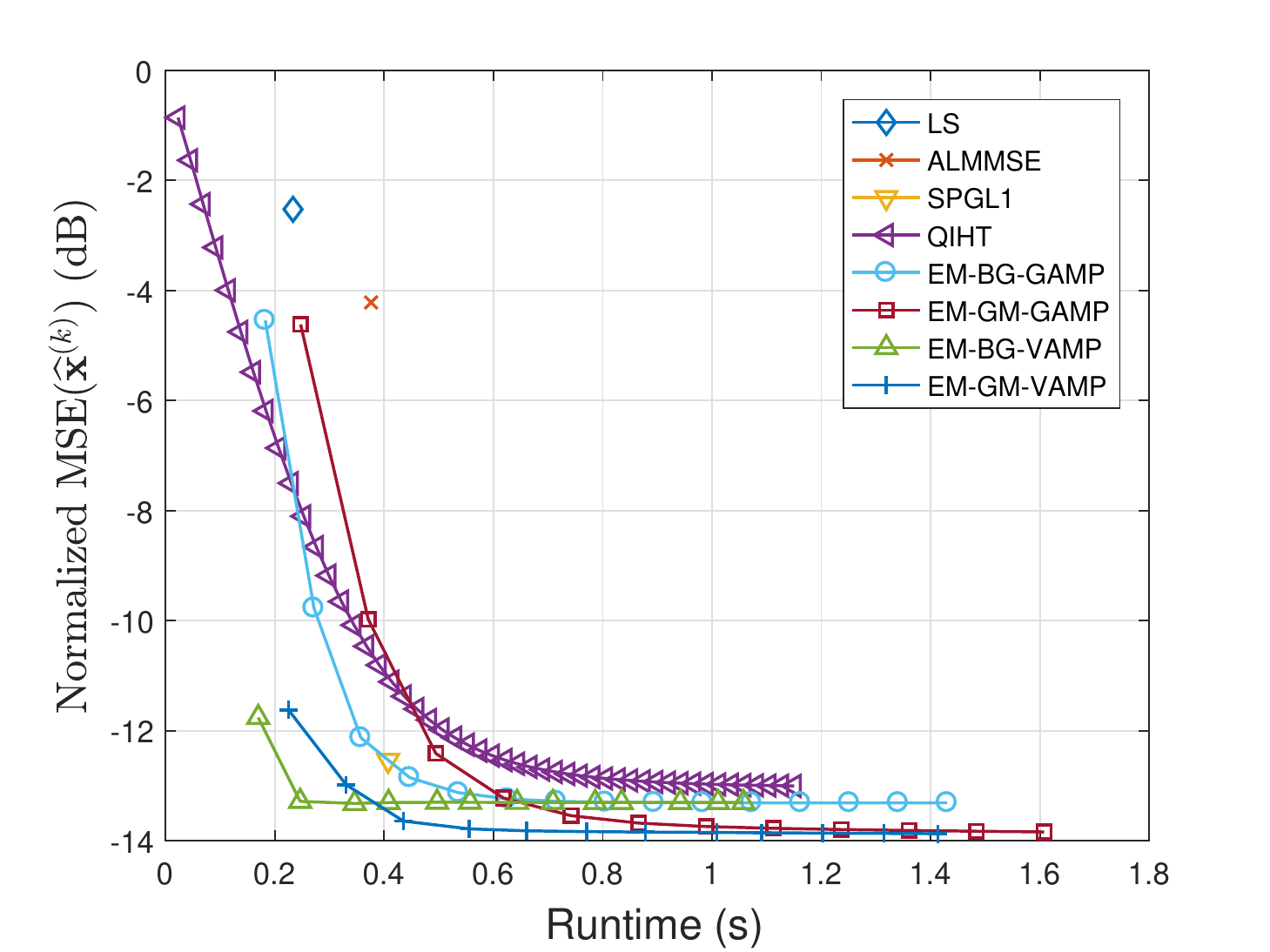}
				\vspace{-0.1cm}
				\centering
				\caption{NMSE versus runtime for several algorithms, where the number of algorithm iterations was varied to obtain different runtimes.  Here, SNR $= 0$~dB, the ADC had 4 bits, and the training was shifted ZC of length $\Np=2048$.}\label{fig:NMSE_vs_Runtime}
			\end{centering}
			\vspace{-0.3cm}
		\end{figure}

	To further investigate algorithm complexity, we plot NMSE versus runtime in \figref{fig:NMSE_vs_Runtime}.  We controlled the runtime of the QIHT, EM-GAMP and EM-VAMP algorithms by varying the number of iterations. The other three (non-iterative) algorithms are each represented by a single point on the plot.
	The figure shows that EM-GM-VAMP gives the best NMSE-complexity trade-off for runtimes $>0.4$~s,
    while EM-BG-VAMP gives the best trade-off for runtimes between $0.17$~s and $0.4$~s.
    The relatively long time required to complete the first iteration of EM-VAMP is likely due to the object-oriented MATLAB implementation\footnote{MATLAB codes for EM-GAMP and EM-VAMP are available from \url{http://sourceforge.net/projects/gampmatlab/}.} and would likely not be an issue in a dedicated implementation.
    Although QIHT can complete a few of its iterations before the first EM-BG-VAMP iteration, the corresponding estimates are probably not useful because their NMSE is so poor.
        The figure also shows that the NMSE-complexity frontier of EM-BG-GAMP is not too far away from that of EM-GM-VAMP, and that SPGL1 achieves one specific point on the EM-BG-GAMP frontier.
        Finally, the figure shows that the LS and ALMMSE algorithms are far from optimal.

\revision{Table~\ref{tab:Complexity} summarizes the complexity scalings of the algorithms under test.
        The complexities of the LS and ALMMSE approaches are dominated by the inversion of an $\Np\times\Np$ matrix, where $\Np$ is an integer multiple of $\Nt L$, and so their complexities scale as $\mathcal{O}(\Nt^3 L^3)$.
        The complexities of the EM-AMP, SPGL1, and QIHT algorithms are dominated by the matrix-vector multiplications with $\bA$ and $\bA^*$.  Matrix $\bA$ is of size $\Nx \times \Ny$, where $\Nx=\Nt\Nr L$ and $\Ny=\Nr\Np$.  Since $\Np$ is an integer multiple of $\Nt L$, both $\Nx$ and $\Ny$ are $\mc{O}(\Nt\Nr L)$.  And since we use an FFT-like algorithm to implement the matrix-vector multiplies with $\bA$ and $\bA^*$, the complexity order of EM-AMP, SPGL1, and QIHT scales as $\mathcal{O}\big(\Nt \Nr L \log(\Nt \Nr L)\big)$.
        In our simulations, $\Nt$ and $\Nr$ are of a similar size, and so we conclude that the complexity scaling of EM-AMP, SPGL1, and QIHT is much lower than that of LS and ALMMSE.
           \begin{table*}
           	\centering
           	\caption{Algorithmic Complexity}
           	\label{tab:Complexity}
           \begin{tabular}{|c|c|}
           	\hline
           	Algorithm &  Complexity Scaling\\
           	\hline
           	LS & $\mathcal{O}(\Nt^3 L^3)$ \\
           	\hline
           	ALMMSE &  $\mathcal{O}(\Nt^3 L^3)$\\
           	\hline
           	SPGL1 & $\mathcal{O}\left(\Nt \Nr L \log\left(\Nt \Nr L\right)\right) $ \\
           	\hline
           	QIHT & $\mathcal{O}\left(\Nt \Nr L \log\left(\Nt \Nr L\right)\right) $ \\
           	\hline
           	EM-AMP &  $\mathcal{O}\left(\Nt \Nr L \log\left(\Nt \Nr L\right)\right) $\\
           	\hline
           \end{tabular}
        \end{table*}
		}
	
	\subsection{Effect of SNR, ADC resolution, and Training Length}
\begin{figure}[t]
			\begin{centering}
				\includegraphics[width=0.95\columnwidth]{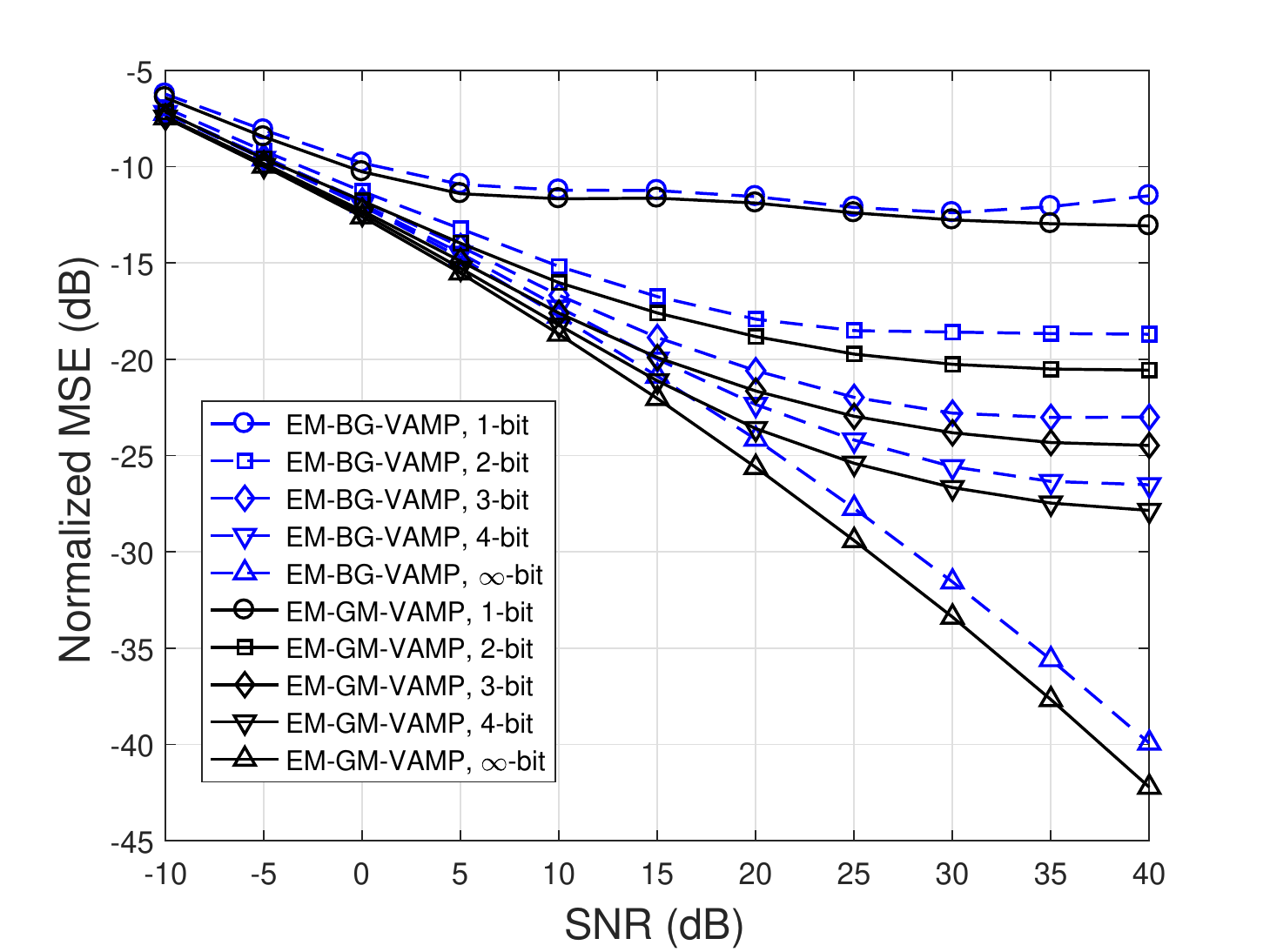}
				\vspace{-0.1cm}
				\centering
				\caption{NMSE versus SNR for the EM-VAMP algorithm under various ADC resolutions. Here, the training was shifted ZC with length $\Np=2048$.}\label{fig:NMSE_SNR_BG_GM_2048}
			\end{centering}
			\vspace{-0.3cm}
		\end{figure}

	We now investigate estimation performance versus SNR, ADC resolution, and training length.
	\figref{fig:NMSE_SNR_BG_GM_2048} shows the NMSE of the VAMP algorithms versus SNR.
	The figure shows that, at low SNR ($< 0$ dB), the performance gap between 1-bit ADC and infinite-bit ADC is only $2$~dB.  But as the SNR increases, the gap between 1-bit and infinite-bit performance grows. Thus, higher resolution ADCs provide significant benefits only at higher SNRs. To reduce power consumption and cost, few-bit ADCs should be deployed when the SNR is low.
		
\begin{figure}[t]
			\begin{centering}
				\includegraphics[width=0.95\columnwidth]{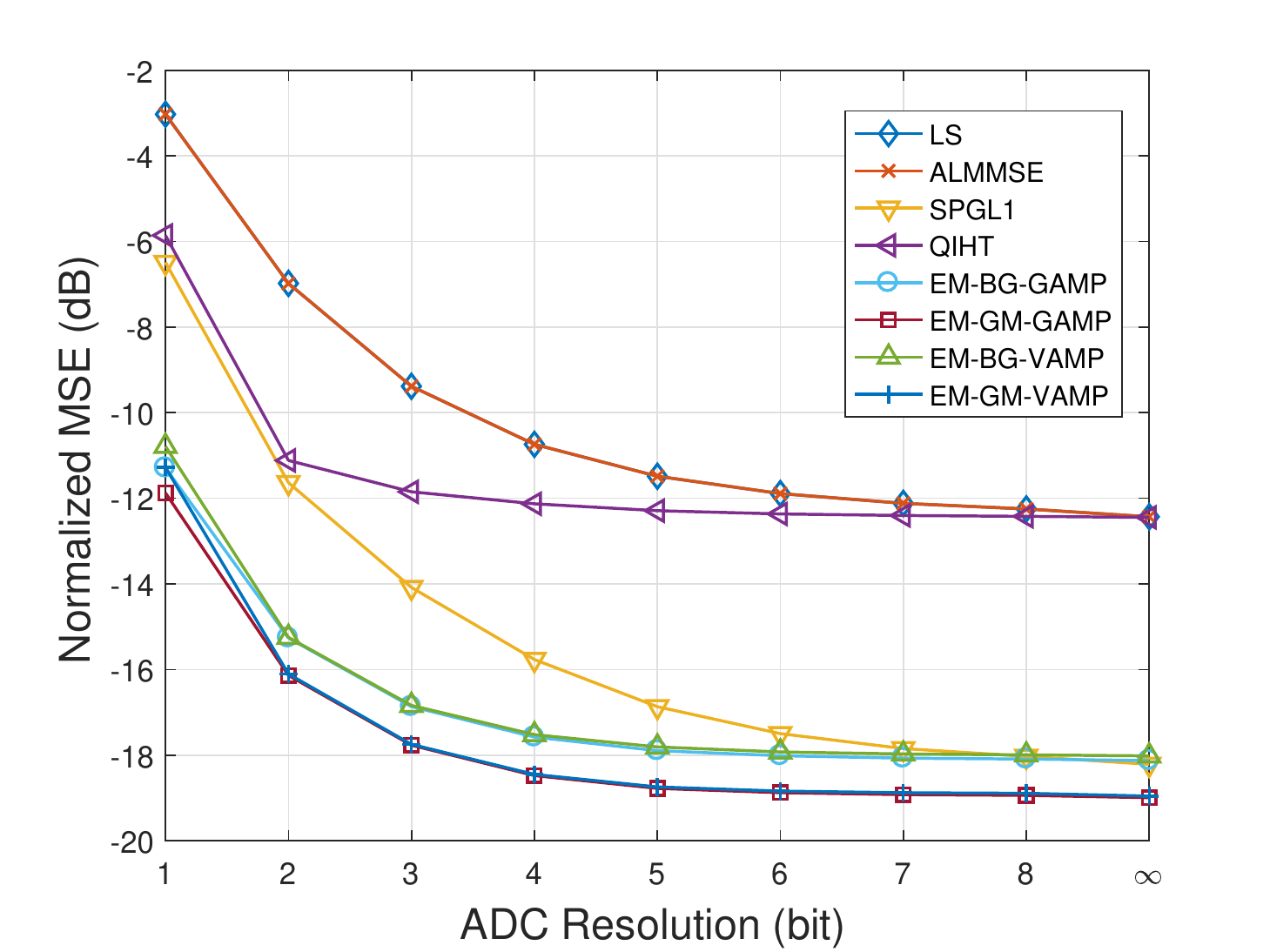}
				\vspace{-0.1cm}
				\centering
				\caption{NMSE versus ADC resolution for several algorithms. Here, SNR $= 10$~dB and the training was shifted ZC with length $\Np=2048$.}\label{fig:NMSE_bit_10dB}
			\end{centering}
			\vspace{-0.3cm}
		\end{figure}

	\figref{fig:NMSE_bit_10dB} shows NMSE versus ADC resolution for all algorithms under test when SNR $=10$~dB.  The figure shows that NMSE decreases with ADC resolution for all algorithms, but not in a uniform way.
        For example, SPGL1 and QIHT perform similarly with $1$ and $2$ bits of resolution, but SPGL1 benefits from higher ADC resolution while QIHT does not.
        In fact, SPGL1 performs as good as EM-BG-GAMP/VAMP for $>7$ bits, but significantly worse with few bits, because it does not leverage the structure of the quantization error.
        The figure also shows that, when the EM-GAMP/VAMP are used, there is little benefit in increasing the ADC resolution above $4$ bits at this SNR.

\begin{figure}[t]
			\begin{centering}
				\includegraphics[width=0.95\columnwidth]{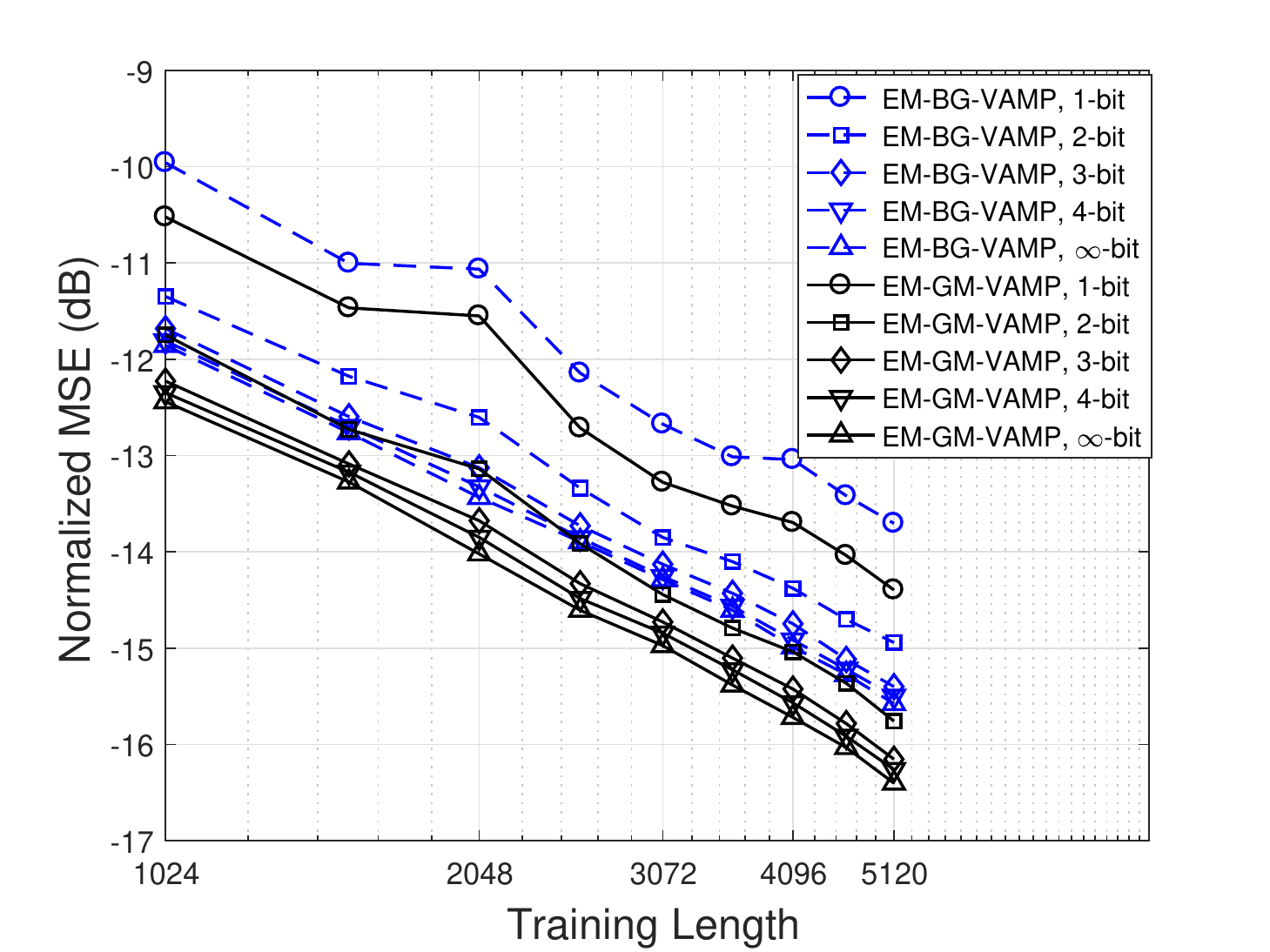}
				\vspace{-0.1cm}
				\centering
				\caption{NMSE versus training length for the EM-VAMP algorithm under various ADC resolutions. Here, the training was shifted ZC and SNR $= 0$~dB. \revision{It is observed that the estimation error exponentially decreases with the training length.}}\label{fig:NMSE_Nb_0dB}
			\end{centering}
			\vspace{-0.3cm}
		\end{figure}

	\figref{fig:NMSE_Nb_0dB} shows the NMSE of the VAMP algorithms versus training length $\Np$.
                We note that, as the training length varies from $1024$ to $5120$, the sampling rate relative to Nyquist, i.e., $\frac{\Ny}{\Nx} = \frac{\Np \Nr}{L \Nt  \Nr} = \frac{\Np}{1024}$, varies from $1$ to $5$.
		The figure shows that the NMSE decays polynomially with the training length.
                In particular, NMSE $\propto \Np^{-\alpha}$ with $\alpha \approx 1/2$.

	\subsection{Mutual Information and Achievable Rate Bounds}
	In this section, we investigate the effect of channel-estimation accuracy on mutual information and achievable rate, which are important metrics for communication systems. Our methodology is inspired by that in \cite{Schniter_Asilomar14}.

        We consider OFDM transmission with $\Nb$ subcarriers. For the $k$th subcarrier, denote the true frequency-domain MIMO channel by $\bG_{k}\in\Complex^{\Nr\times\Nt}$, the transmitted signal by $\bs_k$, and the variance-$\sigma_{w_\mathrm{f}}^2$ additive Gaussian noise by $\bw_k$.
	Using the linearized model \eqref{eq:linearized}, the $k$th subcarrier output is
	\begin{align}
	\by_k = (1- \eta_b) \bG_{k} \bs_k + \hat{\bw}_k,
	\end{align}
        with effective noise $\hat{\bw}_k=(1-\eta_b)\bw_k + \bwqk$.
        Using the Bussgang approximation \eqref{eq:white}, we can approximate its covariance (conditional on $\{\bG_l\}_{l=1}^{\Nb}$) by
	\begin{align}
        \mathbb{E} \left[\hat{\bw}_k \hat{\bw}_k^* \,|\, \{\bG_l\}\right]
        &\approx (1-\eta_b) \Bigg[ \sigma_{w_\mathrm{f}}^2\bI_{\Nr}
                \\&\quad
           + \eta_b\mr{diag}\left( \frac{1}{\Nb}\sum_{l=1}^{\Nb}\bG_l\bR_l\bG_l^* \right) \Bigg] \triangleq \hat{\bSigma}_{\hat{w}_k} ,
                \nonumber
	\end{align}
        where $\bR_k \triangleq \mathbb{E}[\bs_k \bs_k^*]$ and the averaging over $\Nb$ subcarriers occurs because quantization is performed in the time-domain.

	
	Suppose that $\bs_k$ is precoded based on the estimated frequency-domain channel $\hat{\bG}_k$, which is computed from the estimated time-domain channel $\{\hat{\bH}[\ell]\}_{\ell=0}^{L-1}$.
	In particular,
	\begin{align}
	\bs_k = \sum_{m=1}^{\min(\Nt,\Nr)} \hat{\bv}_{km} \sqrt{p_{km}} d_{km},
	\end{align}
	where $\hat{\bv}_{km}$ is the $m$th right-singular vector of $\hat{\bG}_k$, $p_{km}\geq 0$ are the powers allocated by the waterfilling algorithm, and $d_{km}$ is a unit-variance message-bearing symbol.
	
	If each stream is decoded independently and the effective noise $\hat{\bw}_k$ is treated as if it has the worst-case Gaussian distribution with covariance $\hat{\bSigma}_{\hat{w}_k}$, then the mutual information between $\{\bs_k \}_{k=1}^{\Nb}$ and $\{\by_k\}_{k=1}^{\Nb}$ conditioned on $\{\bG_k\}_{k=1}^{\Nb}$ (in units of bps/Hz) is lower bounded by \cite{Tse_05}
	\begin{align} \label{eq:mutual_information}
	\lefteqn{ \underline{I}\left(\{\bs_k \}_{k=1}^{\Nb}, \{\by_k\}_{k=1}^{\Nb} \left| \{\bG_k\}_{k=1}^{\Nb} \right.\right) }\nonumber \\
        & \triangleq \mathbb{E}\left\{
        \frac{1}{\Nb} \sum_{k=1}^{\Nb} \sum_{m=1}^{\min(\Nt,\Nr)}
        \log_2 \Bigg(1
                \right. \\ & \quad \left. +
        \frac{(1 - \eta_b)^2
              |\hat{\bu}_{km}^* \bG_k \hat{\bv}_{km}|^2 p_{km}}
             {\hat{\bu}_{km}^* \hat{\bSigma}_{\hat{w}_k} \hat{\bu}_{km}
              + \sum_{n\neq m}(1-\eta_b)^2 |\hat{\bu}_{km}^* \bG_k \hat{\bv}_{kn} |^2 p_{kn} }
        \Bigg) \right\} \nonumber
	\end{align}
	where $\hat{\bu}_{km}$ is the $m$th left-singular vector of $\hat{\bG}_k$.
        The expectation in \eqref{eq:mutual_information} is taken over $\{\bG_k\}_{k=1}^{\Nb}$ and the effective noise $\{\hat{\bw}_k\}_{k=1}^{\Nb}$.
        (Because the channel estimates depend on $\bG_k$ and $\hat{\bw}_k$, so do $\hat{\bu}_{km}$ and $\hat{\bv}_{km}$.)
	If the coherence time of the channel is $\Nco$ symbols, then a lower bound on the achievable rate (in bps/Hz) is given by \cite{Tse_05}
	\begin{align}
	R \geq \frac{\Nco-\Np}{\Nco} \underline{I}\left( \{\bs_k\}_{k=1}^{\Nb}; \{\by_k\}_{k=1}^{\Nb} \Big| \{\bG_k\}_{k=1}^{\Nb} \right) ,
        \label{eq:achievable_rate}
	\end{align}
	where $\frac{\Nco-\Np}{\Nco}$ represents loss due to training overhead.
\black

\begin{figure}[t]
			\begin{centering}
				\includegraphics[width=0.95\columnwidth]{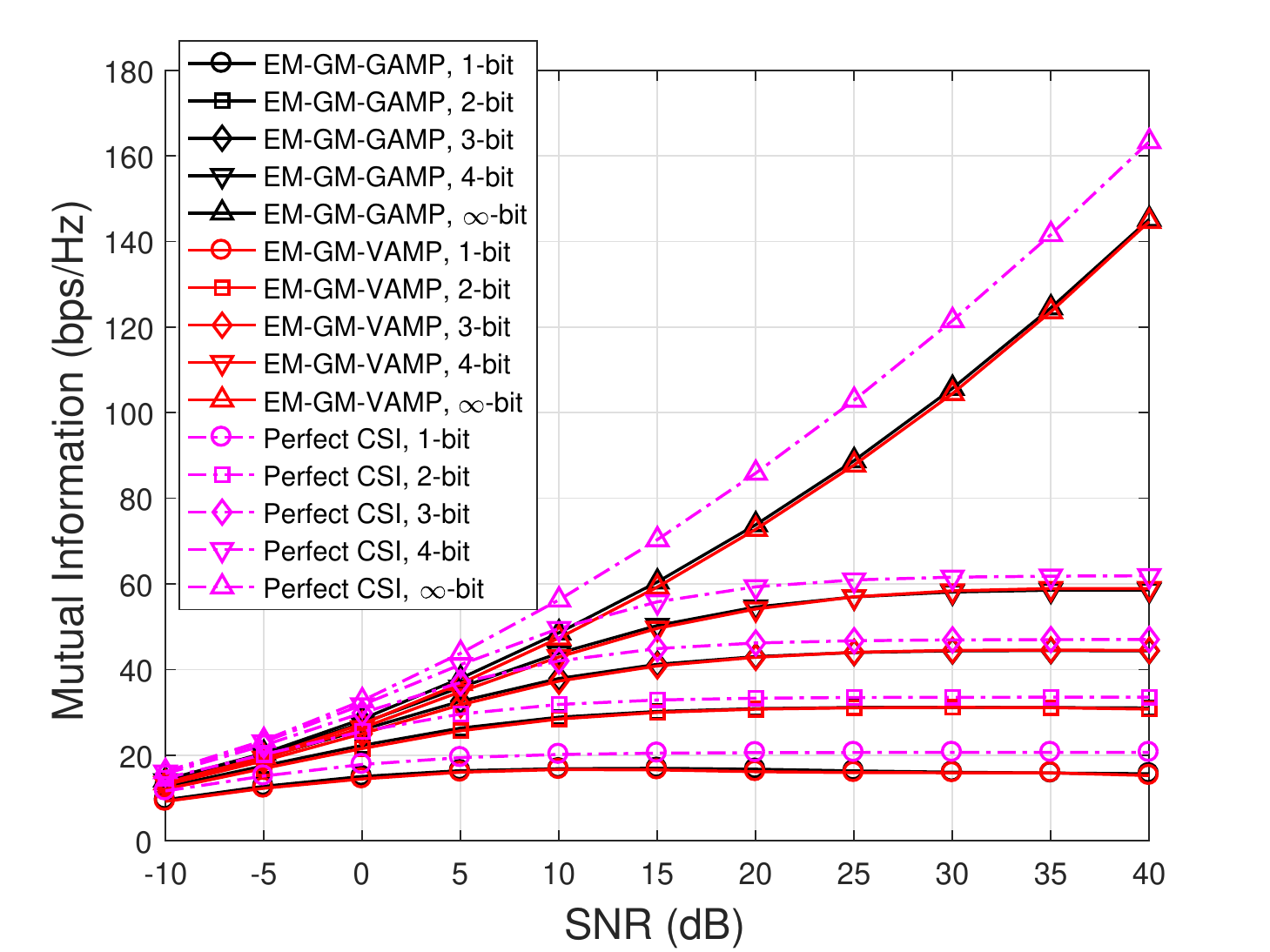}
				\vspace{-0.1cm}
				\centering
				\caption{Mutual information lower bound \eqref{eq:mutual_information} versus SNR for EM-VAMP under various ADC resolutions. Here, the training was shifted ZC with length $\Np=2048$.}\label{fig:MI_vs_SNR}
			\end{centering}
			\vspace{-0.3cm}
		\end{figure}

	\figref{fig:MI_vs_SNR} plots the mutual information lower bound \eqref{eq:mutual_information} versus SNR under EM-GAMP/VAMP channel estimates and perfect CSI at various ADC resolutions, using Monte-Carlo to approximate the expectation.  The figure shows that using EM-GAMP/VAMP channel estimates results in a relatively small loss in mutual information compared to perfect CSI.
        It also shows that, at low SNR, the mutual information loss of few-bit ADC relative to infinite-bit ADC is small.

\begin{figure}[t]
			\begin{centering}
				\includegraphics[width=0.95\columnwidth]{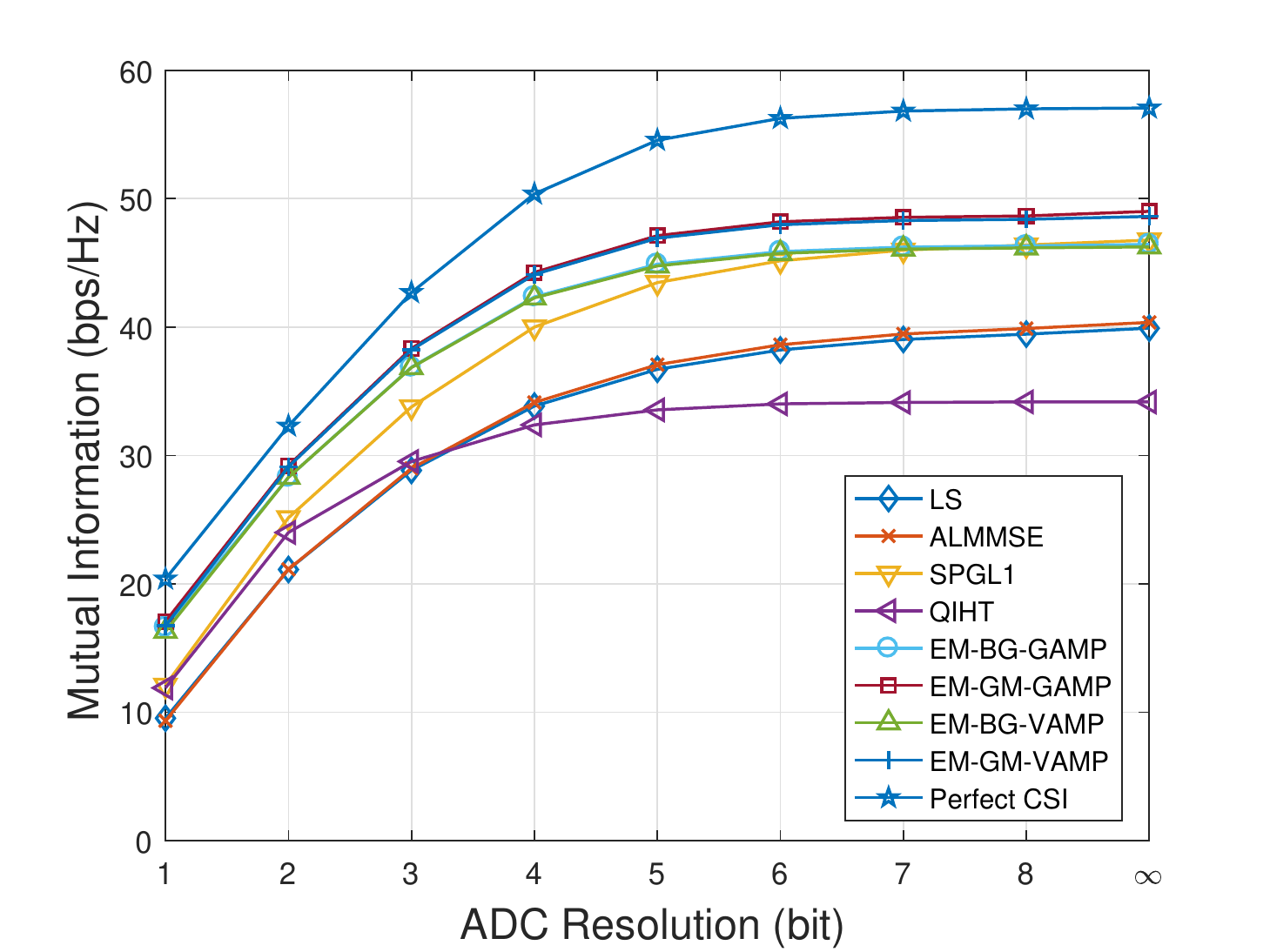}
				\vspace{-0.1cm}
				\centering
				\caption{Mutual information lower bound \eqref{eq:mutual_information} versus ADC resolution for several algorithms. Here, SNR $= 10$~dB and the training was shifted ZC with length $\Np=2048$.}\label{fig:MI_vs_bit}
			\end{centering}
			\vspace{-0.3cm}
		\end{figure}

	\figref{fig:MI_vs_bit} plots the mutual information lower bound \eqref{eq:mutual_information} versus ADC resolution for all the channel-estimation algorithms under test at SNR $=10$~dB. The figure shows that the EM-GM-GAMP/VAMP algorithms achieve the highest mutual information, with EM-BG-GAMP/VAMP close behind at low ADC resolutions.
	The figure also shows that the mutual information saturates when the ADC resolution is $>5$ bits at this SNR.
	
\begin{figure}[t]
			\begin{centering}
				\includegraphics[width=0.95\columnwidth]{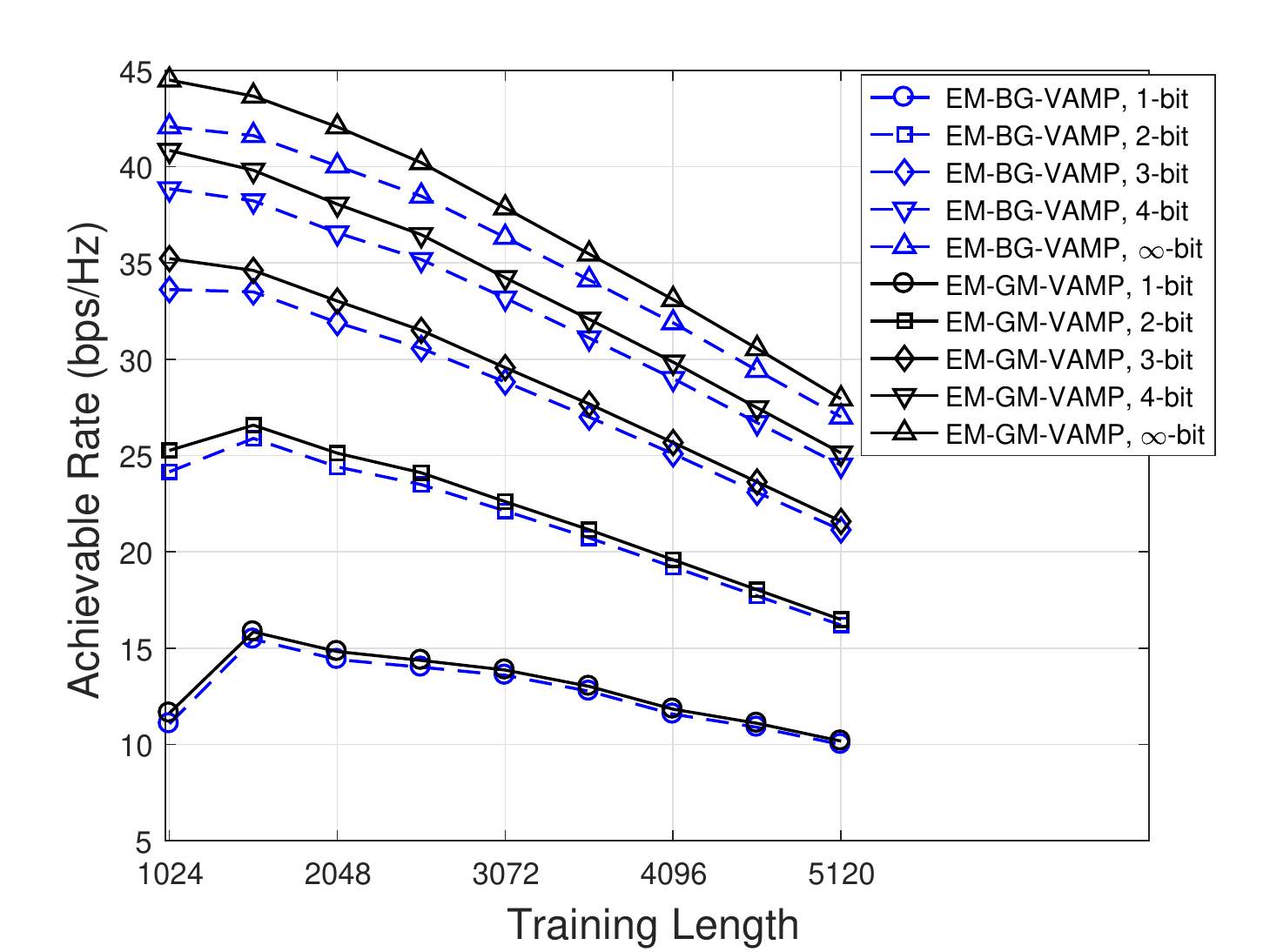}
				\vspace{-0.1cm}
				\centering
				\caption{Achievable rate lower bound \eqref{eq:achievable_rate} versus training length $\Np$ for the EM-VAMP algorithm under various ADC resolutions. Here, SNR $= 10$~dB, the coherence time was $\Nco=10240$, and the training was shifted ZC.}\label{fig:Rate_vs_Np_VAMP}
			\end{centering}
			\vspace{-0.3cm}
		\end{figure}

	\figref{fig:Rate_vs_Np_VAMP} shows the achievable rate lower bound \eqref{eq:achievable_rate} versus training length $\Np$ for the EM-VAMP algorithms. For this experiment, the channel coherence time was assumed to be $\Nco=10240$ symbols. The figure shows that, under these conditions, the optimal training length is $\Np=1536$ when the ADC has $1$ or $2$ bits of resolution.  When the ADC resolution increases to $3$ or $4$ bits, a shorter training length (e.g., $\Np=1024$) is preferred because the cost of each training symbol (in bps/Hz) is larger.
	
\begin{figure}[t]
			\begin{centering}
				\includegraphics[width=0.95\columnwidth]{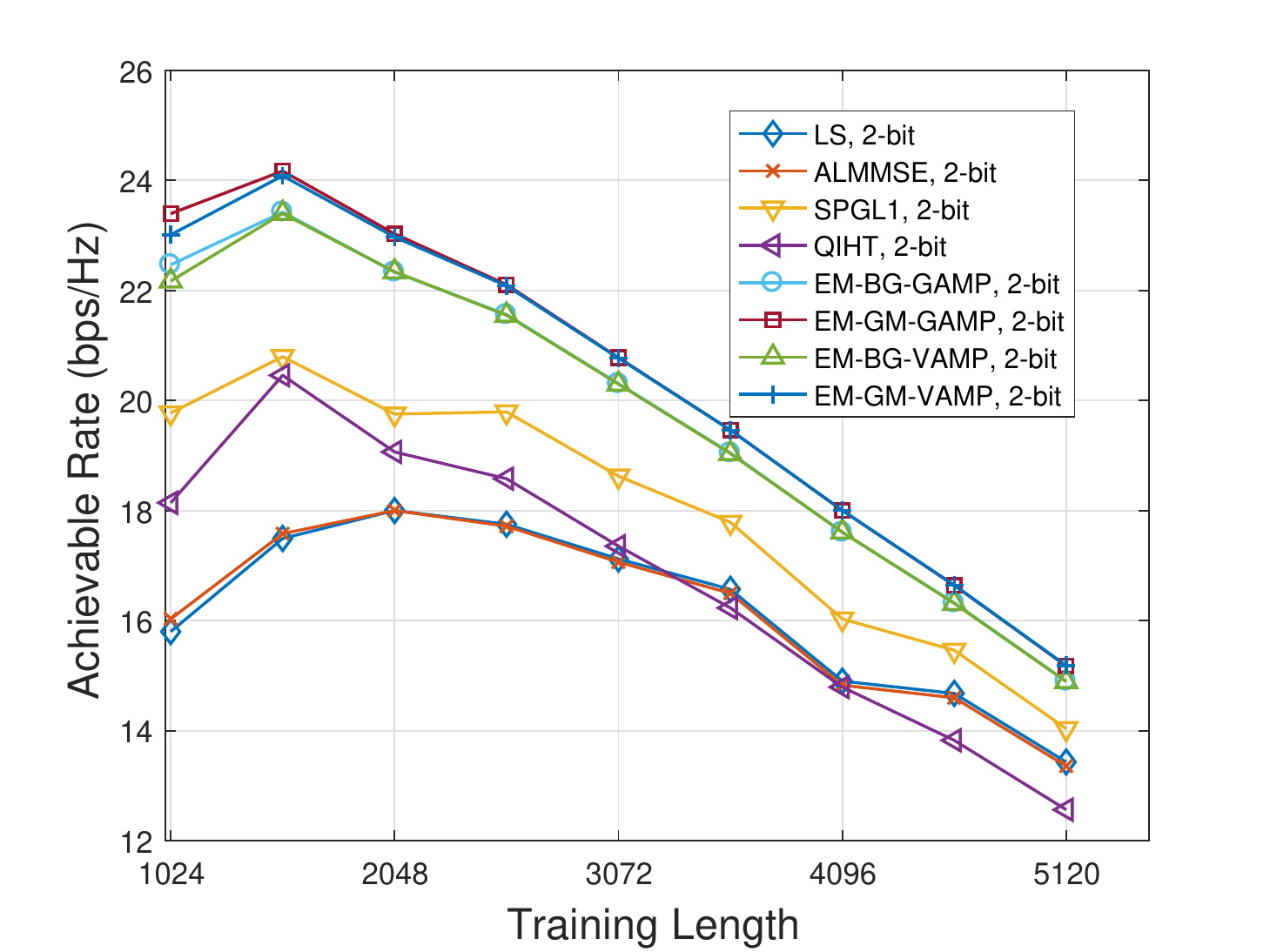}
				\vspace{-0.1cm}
				\centering
				\caption{Achievable rate lower bound \eqref{eq:achievable_rate} versus training length for several algorithms under $2$-bit ADC. Here, SNR $= 10$~dB, the coherence time was $\Nco=10240$, and the training was shifted ZC.}\label{fig:Rate_vs_Np}
			\end{centering}
			\vspace{-0.3cm}
		\end{figure}

	\figref{fig:Rate_vs_Np} shows the achievable rate lower bound \eqref{eq:achievable_rate} versus the training length $\Np$ for all the channel estimation algorithms under test. For this experiment, the ADC has 2-bit resolution. The EM-GM-GAMP and EM-GM-VAMP algorithms provide highest achievable rate and their optimal training length is 1536.
	Note that, for the LS and ALMMSE algorithms, the optimal training length is 2048, which is longer than the other algorithms.
	
	\revision{According to Figs.~\ref{fig:Rate_vs_Np_VAMP} and \ref{fig:Rate_vs_Np}, the training overhead is around 10\%-20\% when the channel coherence length is 10240. The overhead percentage is thus comparable to that used in \cite{Li_Yongzhi_TSP17} for sub-6 GHz massive MIMO channel estimation with a flat fading channel \cite{Li_Yongzhi_TSP17}. Our channel, however, is frequency-selective fading and thus more difficult to estimate.  In addition, a training duration of 1024-2048 symbols seems appropriate for mmWave broadband communication.  Comparing to the 802.11ad standard, for example, if $8 \times 8$ UPA arrays are used with the standard DFT-based codebook, then 128 pilot frames, each consisting of 26 bytes, would be transmitted \cite{Nitsche_COMM14}. The total training length in 802.11ad is thus much longer than what we propose.}

	\section{Conclusion}
		In this paper, we propose a methodology for estimation of broadband mmWave MIMO channels at receivers with few-bit ADCs. The broadband mmWave MIMO channel is sparse in both angle and delay domains, making it natural to apply compressed sensing techniques.  We propose to use computationally efficient AMP algorithms (i.e., EM-GAMP and EM-VAMP) that accurately estimate the channel in the absence of prior information about its distribution (e.g., sparsity), and we enhance those methods with separate channel-norm estimation.  We also design a training scheme, based on shifted ZC sequences, that leads to accurate and computationally efficient estimation with minimal transmitter PAPR.  Finally, we report the results of an extensive simulation study that tests various algorithms, ADC precisions, training sequences, training lengths, SNRs, and runtime limits.  Our simulations investigated channel-estimation MSE as well as mutual information and achievable rate bounds.
		
		From the results of our study, we draw several conclusions.
        First, it is important to exploit the available joint angle-delay domain sparsity in channel estimation.
        Second, the ADC precision should be chosen based on the SNR; at low SNR, the use of few-bit ADCs results in very small loss (in MSE or achievable rate) compared to infinite-bit ADCs.
        Third, the training length should be chosen based on the ADC precision; at lower ADC precision, the achievable rate is maximized by a longer training sequence.
		
		In this paper, the correlation among angle-delay channel coefficients was neglected. A possible direction to improve the channel estimation accuracy is to exploit this correlation.
	\bibliographystyle{IEEEtran}
	\bibliography{IEEEabrv,One_bit_Quantization}
\end{document}